\documentclass[lettersize,journal]{IEEEtran}
\usepackage{amsmath,amsfonts}
\usepackage{array}
\usepackage{textcomp}
\usepackage{stfloats}
\usepackage{url}
\usepackage{verbatim}
\usepackage{graphicx}
\usepackage{cite}
\usepackage{tikz}
\usepackage[normalem]{ulem}
\usepackage{multicol}
\usepackage{graphicx}
\usepackage{pifont}
\usepackage[noend]{algorithmic}
\usepackage[ruled,vlined,linesnumbered,algo2e]{algorithm2e}
\usepackage{enumitem}
\usepackage{multirow}
\usepackage{xcolor}
\usepackage{color}
\usepackage{listings}
\usepackage{float}
\usepackage{ragged2e}
\usepackage{xspace}
\usepackage{soul}
\usepackage{lipsum}
\usepackage{makecell}
\usepackage{subcaption}
\usepackage{fancyhdr}
\usepackage[normalem]{ulem}
\usepackage{url}
\usepackage{flushend}
\usepackage{xcolor}
\usepackage{listings}
\usepackage{multirow}
\usepackage{wrapfig}
\usepackage{etoolbox}
\usepackage{fancyhdr}
\usepackage{wrapfig}

\newcommand{\hlc}[2][white]{{%
    \colorlet{foo}{#1}%
    \sethlcolor{foo}\hl{#2}}%
}

\newcommand{\benchmark}[1]{{\texttt{#1}}}
\newcommand{\proj}{\benchmark{TW}\xspace}
\newcommand{\Fig}[1]{Fig.~\ref{#1}}

\newcommand{\Sec}[1]{Sec.~\ref{#1}}

\renewcommand{\paragraph}[1]{\vspace*{0.15cm}\noindent\textbf{#1}\hspace*{.1cm}}

\newcommand*\circled[1]{\tikz[baseline=(char.base)]{
               \node[shape=circle,fill,inner sep=0.6pt] (char) {\textcolor{white}{#1}};}}

\makeatletter
\newcommand{\removelatexerror}{\let\@latex@error\@gobble}
\makeatother

\begin{document}

\title{Accelerating Sparse DNNs Based on Tiled GEMM}

\author{
        
\IEEEauthorblockN{
Cong~Guo\IEEEauthorrefmark{2},
Fengchen~Xue\IEEEauthorrefmark{2},
Jingwen~Leng\IEEEauthorrefmark{2},~\IEEEmembership{Member, IEEE},
Yuxian~Qiu\IEEEauthorrefmark{3},
Yue~Guan\IEEEauthorrefmark{2},
Weihao~Cui\IEEEauthorrefmark{2},
Quan~Chen\IEEEauthorrefmark{2},~\IEEEmembership{Member, IEEE},
Minyi~Guo\IEEEauthorrefmark{2},~\IEEEmembership{Fellow, IEEE} 
}

\IEEEauthorblockA{
\textit{
\IEEEauthorrefmark{2}Shanghai Jiao Tong University,
\IEEEauthorrefmark{3}NVIDIA}
}
\thanks{
${^\S}$Jingwen Leng and Minyi Guo are corresponding authors of this paper.
}
}



\markboth{Journal of \LaTeX\ Class Files,~Vol.~14, No.~8, August~2021}%
{Shell \MakeLowercase{\textit{et al.}}: A Sample Article Using IEEEtran.cls for IEEE Journals}

\IEEEpubid{0000--0000/00\$00.00~\copyright~2021 IEEE}

\maketitle

\begin{abstract}
Network pruning can reduce the computation cost of deep neural network (DNN) models. However, sparse models often produce randomly-distributed weights to maintain accuracy, leading to irregular computations.
Consequently, unstructured sparse models cannot achieve meaningful speedup on commodity hardware built for dense matrix computations. 
Accelerators are usually modified or designed with structured sparsity-optimized architectures for exploiting sparsity. 
For example, the Ampere architecture introduces a sparse tensor core, which adopts the 2:4 sparsity pattern. 

We propose a pruning method that builds upon the insight that matrix multiplication generally breaks the large matrix into multiple smaller tiles for parallel execution. We present the “tile-wise” sparsity pattern, which maintains a structured sparsity pattern at the tile level for efﬁcient execution but allows for irregular pruning at the global scale to maintain high accuracy. 
In addition, the tile-wise sparsity is implemented at the global memory level, and the 2:4 sparsity executes at the register level inside the sparse tensor core. 
We can combine these two patterns into a ``tile-vector-wise'' (\benchmark{TVW}) sparsity pattern to explore more fine-grained sparsity and further accelerate the sparse DNN models.
We evaluate the \benchmark{TVW} on the GPU, achieving averages of {$1.85\times$}, $2.75\times$, and $22.18\times$ speedups over the dense model, block sparsity, and unstructured sparsity. 
\end{abstract}


\section{Introduction}\label{sec:introduction}
\IEEEPARstart{D}{eep} neural network (DNN) models have achieved and even surpassed human-level accuracy in important domains~\cite{dnnvshuman}. 
For instance, transformer-based models~\cite{46201} in natural language processing (NLP) such as BERT~\cite{devlin2018bert} have dominated the accuracy in various NLP tasks. Despite their high accuracies, DNN models also have significant computational cost, both in training and inference.
The inference latency of modern DNN models could also be excessively high due to the enormous computation cost and memory usage.

One particularly effective and promising approach to reduce the DNN latency is pruning~\cite{han2015learning}, which exploits the inherent redundancy in the DNN models to transform the original, dense model to a sparse model by iteratively removing ``unimportant'' weight elements and retraining the model to recover its accuracy loss. In the end, the sparse model has fewer parameters and, theoretically, less computation cost.

The primary challenge in network pruning is how to balance the model accuracy and execution efficiency.
Such a balance is fundamentally affected by the 
\textit{sparsity pattern} that a pruning approach enforces. Intuitively, a stronger constraint on the sparsity pattern forces certain weights to be pruned and, thus, leads to lower accuracy, and vice-versa.
The most fine-grained pruning approach leads to the so-called element-wise (\benchmark{EW}) sparsity pattern, which prunes weight elements individually and independently, solely by their importance scores~\cite{han2015learning}.
In other words, \benchmark{EW} imposes no constraint on the sparsity pattern and can remove any weight element, leading to  the minimal model accuracy degradation.
However, the pruned sparse model also introduces irregular memory accesses that are unfriendly on commodity architectures.
As a result, \benchmark{EW}-based sparse DNN models usually run slower than the unpruned dense models on these architectures~\cite{hill2017deftnn}.

To realize the acceleration potential of sparse DNN models, researchers have proposed to co-design the sparsity pattern with hardware support.
For instance, many architects have proposed various specialized accelerator designs~\cite{wang2021dual} to exploit the zeros in the aforementioned \benchmark{EW} pattern for latency reduction.
Similarly, prior work proposes the vector-wise (\benchmark{VW}) pattern~\cite{yao2019balanced} that divides a weight column to groups and prunes the same number of elements in a group. 
This sparsity pattern requires the new hardware or the modification of existing hardware~\cite{zhu2019sparse}.
Recently, NVIDIA has published GPU A100 with the Ampere architecture~\cite{a100}, which includes the sparse tensor core with the 2:4 (2-out-of-4) vector-wise sparsity.
In summary, these approaches lead to sparse memory accesses and computation patterns that require hardware support.

\IEEEpubidadjcol

In this work, we propose a novel algorithm that accelerate sparse DNN models on commodity DNN accelerators without hardware modification. Our key observation is that virtually all of today's DNN accelerators implement dense general matrix multiplication (GEMM)~\cite{chellapilla2006high} operations. GEMM-based accelerators~\cite{volta2017whitepaper, jouppi2017datacenter, guo2020balancing, qin2020sigma} are dominant owing to their wide applicability: convolution operations that dominate computer vision models are lowered to the GEMM operation, and NLP models are naturally equivalent to the GEMM operation. Examples include NVIDIA's tensor core~\cite{volta2017whitepaper} and Google's TPU~\cite{jouppi2017datacenter} mentioned above. We propose a new pruning algorithm, which enforces a particular sparsity pattern on pruned models to directly leverage existing GEMM accelerators without modifying the microarchitectures.

In particular, our work exploits the key insight that the matrix multiplication on existing dense GEMM accelerators adopts the tiling approach, which breaks the large matrix into multiple smaller tiles for parallel execution. We propose a tiling-friendly sparsity pattern called \emph{tile-wise sparsity} (or \benchmark{TW}), which maintains a regular sparsity pattern at the tile level for efficient execution but allows for irregular, arbitrary pruning at a global scale through non-uniform tile sizes to maintain high model accuracies. 
To exploit the \texttt{TW} sparsity, we first divide the entire matrix into multiple tiles as in conventional tiled GEMM. We then prune the \textit{entire rows or columns} of each tile according to the collective importance scores of each row and column.
In our sparsity pattern, the tile size dictates the trade-off between model accuracy and execution efficiency.
At one extreme where the tile size equals one, our \texttt{TW} sparsity is equivalent to the \texttt{EW} sparsity.
At the other extreme where the tile size is the same as the matrix size, \texttt{TW} is equivalent to the global structural pruning that prunes rows or columns~\cite{hill2017deftnn}.

Building on top of \texttt{TW}, we further propose hybrid sparsity patterns.
We can overlay the most fine-grained \benchmark{EW} sparsity pattern on top of the \texttt{TW} sparsity and propose \texttt{TEW} sparsity.
With a small fraction of \benchmark{EW} ({e.g., 1.5\%}), the \texttt{TEW} pattern significantly improves the accuracies of the \texttt{TW}-only sparse models. 
For the latest GPU architecture, \texttt{TW} is conducted at the global memory level, and the sparse tensor core of Ampere GPU~\cite{a100} exploits 2:4 \benchmark{VW} sparsity at the register level. Therefore, \texttt{TW} sparsity pattern is orthogonal to the \benchmark{VW} sparsity pattern of the sparse tensor core. We can fuse the \benchmark{TW} and \benchmark{VW} and combine their advantages to build a more fine-grained pattern \benchmark{TVW} and further accelerate the DNN models.
We propose a pruning algorithm that iteratively shapes the weight matrix to meet our hybrid sparsity pattern constraint. Critically, our pruning algorithm dynamically allocates the sparsity budget to each layer to exploit the inherently uneven sparsity distribution across layers.

To maximize the algorithmic benefits of \texttt{TW}/\texttt{TVW}, we provide an efficient software implementation on commodity GPU hardware. 
Two key roadblocks arise as a result of the \texttt{TW} sparsity. First, \texttt{TW} naturally introduces frequently uncoalesced memory accesses due to the pruning pattern. Second, different tiles in \texttt{TW} could have different compute demands due to the different pruning degrees across tiles, which leads to load imbalance and GPU resource under-utilization. We address these challenges through a combination of intelligent data layout and concurrency/batching optimizations.
\proj{} and \texttt{TVW} achieve averages of {$1.85\times$} and {$1.70\times$} latency speedup on the tensor core with only negligible accuracy loss (1\%-3\%). 

The contribution of our work is as follows:
\begin{itemize}[leftmargin=*]
    \item \paragraph{Pattern Design.}  We propose a tile-wise (\benchmark{TW}) sparsity pattern  to balance the model accuracy and execution efficiency on the existing dense accelerator.
    To further exploit the new sparse architecture, we propose the {tile-vector-wise} (\benchmark{TVW}) sparsity which can fuse \benchmark{TW} with the \benchmark{VW} to achieve a more fine-grained sparsity pattern for the DNN pruning algorithm.
    \item \paragraph{Pruning Algorithm.} We propose a multi-stage pruning algorithm that gradually shapes the weight matrix to \benchmark{TW} pattern and dynamically allocates the sparsity budget at the layer level to overcome the uneven sparsity distribution.
    \item \paragraph{Implementation.} We provide an efficient implementation of tile sparsity on commodity GPUs equipped with tensor core, and demonstrate significant speedups on state-of-the-art DNN models.
    We further optimize the implementation of \benchmark{TW} with the compressed tile offset (CTO) to execute the \benchmark{TW} in a single CUDA kernel.
    \item \paragraph{New Hardware Support.} \benchmark{TVW} sparsity can be implemented on the sparse tensor core of the newest Ampere GPU architecture~\cite{a100} to achieve better accuracy at a high sparsity level and further accelerate DNN models.
\end{itemize}


\section{Background and Related Work}
\label{sec:background}
We summarize the recent efforts to reduce those models' execution latency, which include building specialized hardware accelerators and applying algorithmic pruning optimization to reduce the size and computation cost of DNNs.

\subsection{Hardware Acceleration}\label{background:execution}

We explain the computation characteristics of these models and common optimizations to reduce their execution latency.

\paragraph{Dense Model.}
General matrix multiplication (GEMM) is a critical computation in the original dense DNN models.
The fully connected layer and LSTM layer are native GEMM operations, while the convolutional layer can be converted to GEMM through the \benchmark{img2col} transformation~\cite{chetlur2014cudnn, zhou2021characterizing}.
The attention heads in BERT can also be computed with GEMM operations, and the computation of multiple heads could be combined into one large GEMM. 

\begin{figure}[t]
    \begin{center}
    \includegraphics[width=0.45\textwidth]{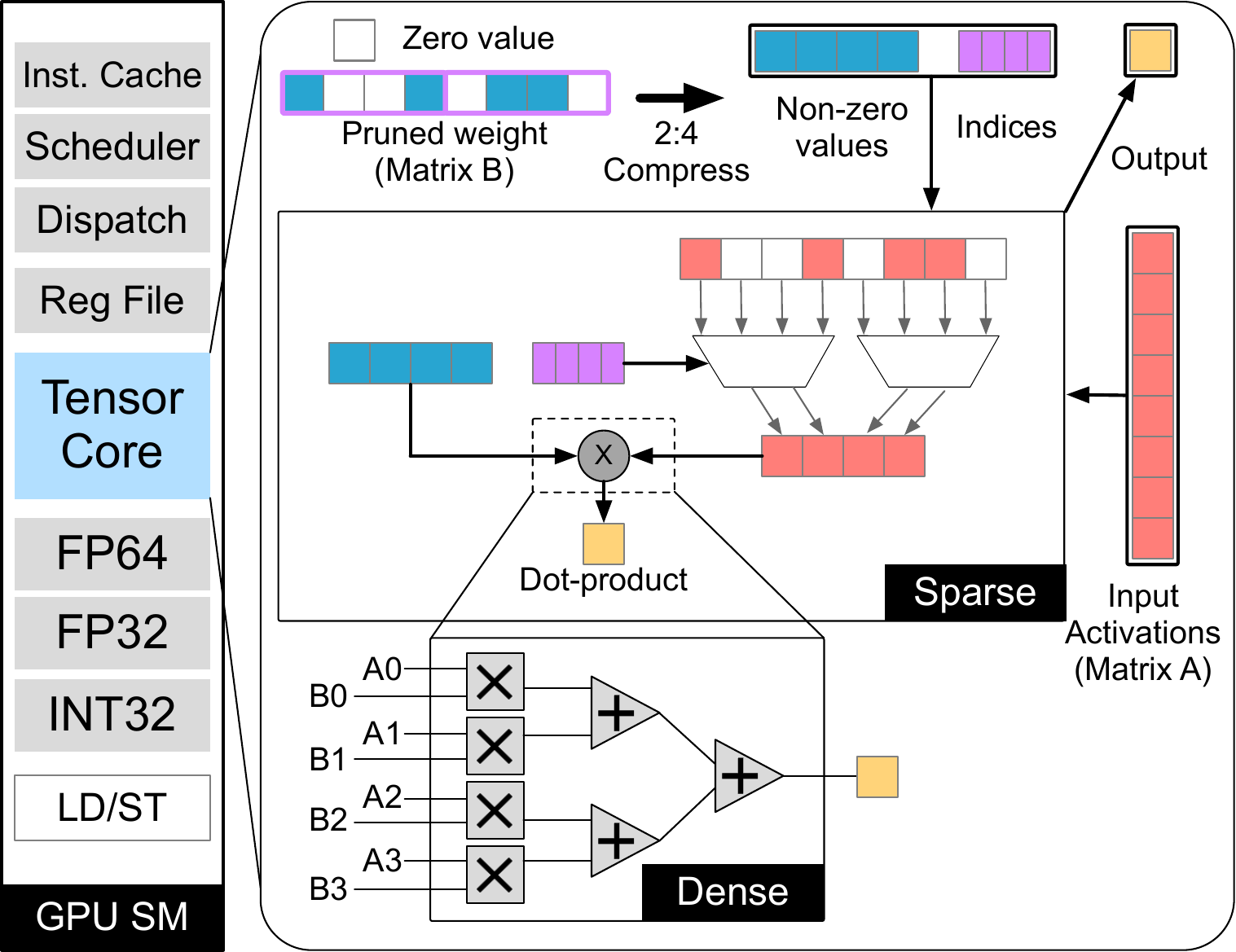}
    \caption{The Ampere GPU architecture~\cite{a100} introduces the sparse tensor core with 2:4 (2-out-of-4) vector-wise sparsity based on the dense tensor core.}
    \label{fig:stc}
    \end{center}
    \vspace*{-0.5cm}
\end{figure}

\noindent\textbf{GEMM Acceleration.}
To reduce the model execution latency, NVIDIA adds tensor cores on the GPU since Volta architecture~\cite{volta2017whitepaper}, which run a fixed size matrix multiplication.
As shown in \Fig{fig:stc}, the tensor core is essentially an accelerator for the GEMM. To exploit the DNN sparsity, NVIDIA upgrades the tensor core with a new 2:4 vector-wise sparsity feature that delivers a further doubling of throughput in the Ampere architecture~\cite{a100}, as shown in \Fig{fig:stc}.
The tensor core can also reduce the computation precision to FP16, Int8, and Int4 to support DNN quantization~\cite{gupta2015deep, nagel2020up,  guo2022squant, guo2022ant, guo2023olive}, which can also reduce the DNN model size.
Our proposed sparsity is orthogonal to the quantization.
Another example of the GEMM accelerator is TPU~\cite{jouppi2017datacenter} which is based on a $128 \times 128$ systolic array.
The cuDNN~\cite{chetlur2014cudnn} library implements different DNN layers for efficient execution on GPU, where the GEMM computation can use the closed-source cuBLAS library~\cite{nvidia2019toolkit} or open-sourced CUTLASS library~\cite{cutlass2019}.
For large language models, some work~\cite {chen2021re, peng2022length, zhai2023bytetransformer} focused on optimizing the sparsity of self-attention operation with long sequences.

\begin{figure*}[t]
    \begin{center}
    \includegraphics[width=1.0\textwidth]{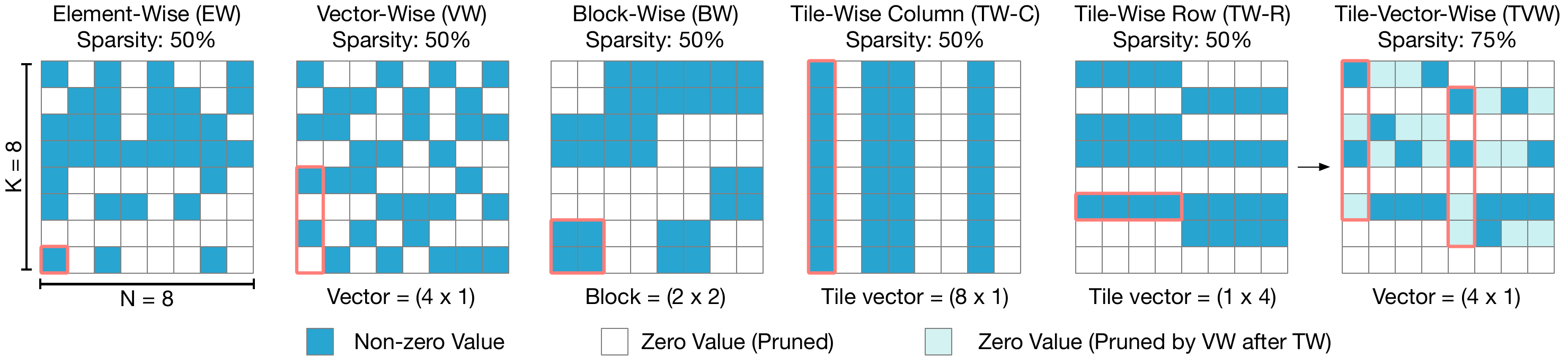}
    \caption{Comparison of six patterns with sparsity.
    }
    \label{fig:pruning_patterns}
    \end{center}
    \vspace*{-0.8cm}
\end{figure*}

\paragraph{Sparse Model.}
Recently, researchers have started to apply pruning~\cite{han2015learning} to DNN models, which exploit the inherent redundancy in the model to transform the original, dense model into a sparse model. In the end, the sparse model has fewer parameters and, theoretically, less computation cost.
Executing sparse models relies on sparse matrix representation such as compressed sparse row (CSR)~\cite{nvidia2019toolkit} and sparse GEMM operations, which are supported on GPU by cuSparse~\cite{nvidia2019toolkit} library.
However, as the GPU is initially designed for dense operations, the speedup of the sparse model over the dense model is usually negative unless the sparsity ratio is very large (over 95\% reported by prior work~\cite{wen2016learning}). 
As such, researchers begin to put various shape constraints on the pruning pattern and also propose to transform the existing architecture to execute those sparse models.
 For example, the recent work proposes new sparsity patterns that need to modify the existing dense GEMM accelerator~\cite{zhu2019sparse, wang2021dual}.

Unlike the prior microarchitecture-centric work, we propose a software-only acceleration of sparse DNN models on the dense GEMM accelerator, like the tensor core.
We exploit the tile execution of GEMM computation and propose a tiling-friendly, \benchmark{tile-wise} sparsity pattern to balance the model accuracy and compatibility for the dense GEMM accelerator.

\subsection{Sparsity Pattern}
Many sparsity patterns~\cite{han2015learning, zhu2019sparse,yao2019balanced, a100, narang2017block} have been proposed to prune the weight tensors. 
\Fig{fig:pruning_patterns} illustrates six sparsity patterns.
For the sparse models, their weight tensors are static and pro-processed before the inference stage. However, for the activation tensors, the methods need to prune on the fly at runtime due to the irregular memory accesses. Therefore, these irregular patterns need unique software design, e.g., CSR representation for the \benchmark{element-wise} pattern~\cite{nvidia2019toolkit}.

\paragraph{\benchmark{Element-wise.}}
\benchmark{Element-wise (EW)} removes the individual weight element solely by its importance score rank. 
For instance, prior work~\cite{han2015learning} proposes to remove weight elements with small magnitude.
This approach imposes no constraints on the sparsity pattern and could remove most of the weights among all pruning methods.
Thus, it is also called unstructured pruning.
However, the randomly distributed non-zero weights lead to substantial irregular memory accesses, which impose great challenges for efficient hardware execution.
As such, researchers propose other two more structured pruning methods.

\paragraph{\benchmark{Vector-wise.}}
The second sparsity pattern shown in the middle of \Fig{fig:pruning_patterns}, \benchmark{vector-wise (VW)}~\cite{zhu2019sparse,yao2019balanced}, divides a column in the weight matrix to multiple vectors.
Within each vector, it prunes a fixed portion of elements by the rank of their importance scores.
This approach preserves the randomness within each vector for model accuracy.
Meanwhile, it maintains the regular structure  for efficient execution, where different vectors have the same number of non-zero weight elements. The GPU A100~\cite{a100} develops the sparse tensor core using the 2:4 \benchmark{VW} sparsity pattern with fixed 50\% sparsity.

\paragraph{\benchmark{Block-wise.}}
\benchmark{Block-wise (BW)}~\cite{narang2017block} divides the weight matrix into small blocks, and treats a block as the basic pruning unit. 
In other words, the \benchmark{EW} sparsity pattern is a special case of  the \benchmark{BW} sparsity pattern, which expands a $1\times 1$ block to an $n \times n$ block.
The structural sparsity pattern \benchmark{BW} leads to the efficient execution of sparse models. 

\paragraph{\benchmark{Tile-wise.}}
\benchmark{Tile-wise (TW)}~\cite{guo2020accelerating} pattern has two types,  i.e., \benchmark{TW-C} and \benchmark{TW-R}, addressing vertical and horizontal dimensions of weight tensor and is based on the tiled GEMM algorithm, which is widely used  in the GEMM accelerator~\cite{nvidia2019toolkit, cutlass2019}. For each tile (sub-matrix), \benchmark{TW} can simultaneously prune the column and row for the weight tensor. We will introduce the details in \Sec{sec:tile_sparsity}.

\paragraph{\benchmark{Tile-vector-wise.}} 
The last pattern is our proposed \benchmark{tile-vector-wise (TVW)}, which is a hybrid sparsity fusing \benchmark{TW} and \benchmark{VW}. We present the \benchmark{TVW} design in \Sec{sec:tile_sparsity}.

\begin{figure*}[t]
    \begin{center}
    \includegraphics[width=\linewidth]{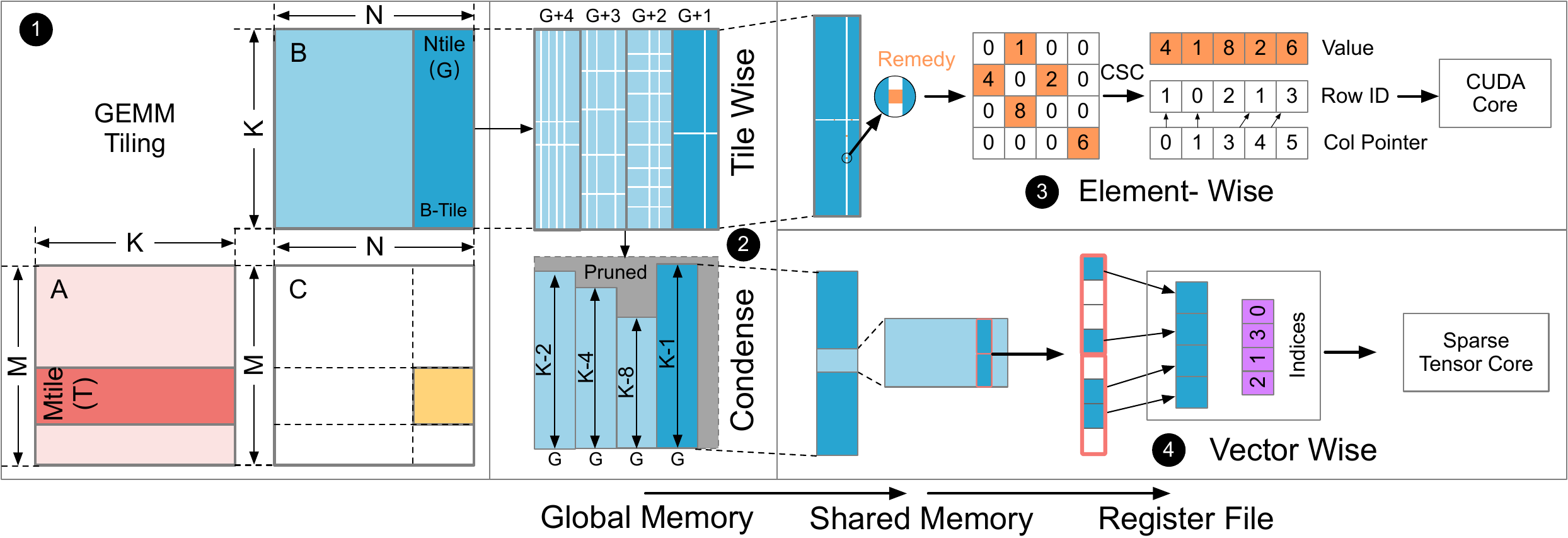}
    \caption{
   The overview of \benchmark{TW} sparsity pattern that exploits the tiled GEMM to maintain the GEMM-compatible execution.
    }
    \label{fig:tw}
    \end{center}
    \vspace*{-0.2cm}
\end{figure*}

\section{Tile-wise and Hybrid Sparsity}
\label{sec:tile_sparsity}

In this section, we present the details of our proposed \benchmark{TW}, \benchmark{TEW}, and \benchmark{TVW} sparsity pattern.
The \benchmark{TW} sparsity pattern leverages the tiled execution of matrix multiplication, originally designed to exploit the parallel computation resources.
The \benchmark{TW} sparsity pattern introduces irregularity in the global matrix. Still, it maintains the computation regularity of individual matrix tiles exploiting the memory condensing approach crossing the shared and global memory.
The \benchmark{VW} sparsity pattern exploits the sparse tensor core to execute the 2:4 (2-out-of-4) \benchmark{VW} sparsity at the register level.
Therefore, these two patterns leverage different levels of GPU memory hierarchy, and \benchmark{TVW} can be implemented simultaneously in a single CUDA kernel.
We also show that the tile-wise sparsity pattern can be overlaid with the most fine-grained element-wise pattern (\benchmark{TEW}) to increase the sparsity of pruned models and reduce their accuracy loss.

\subsection{Tiling and Pruning Co-design}
\label{subsec:tw}

As \Sec{sec:background} explains, the dominant computation in deep neural network models is the general matrix multiplication (GEMM).
In this subsection, we first present the details of tiled matrix multiplication.
We then propose to co-design the matrix tiling and deep neural network pruning, which leads to the \benchmark{tile-wise (TW)} sparsity pattern.
We explain how \benchmark{TW} maintains the compatibility on the dense GEMM accelerator and the composability with the fine-grained sparsity pattern.

\Fig{fig:tw} \circled{1} shows one level tiling of matrix multiplication on the GPU.
The GEMM computes $C = A \times B$ with input matrix $A$ ($M\times K$), weight matrix $B$ ($K \times N$), and output matrix $C$ ($M \times N$). 
Since modern high-performant microprocessors mainly adopt the manycore architecture, the tiled execution of output matrix $C$ breaks the entire GEMM computation into multiple ones such that they can run on multiple cores for parallel execution.
Specifically, each core (or streaming multi-processor, SM in NVIDIA GPU) computes one tile with the size of $T \times G$ ($Mtile \times Ntile$).
Consequently, the core only loads $T$ rows of input matrix $A$ and $G$ columns of weight matrix $B$ (called $B_{tile}$).
With the output matrix tile size of $T \times G$, the $K\times N$ weight matrix $B$ is divided to $\lceil \frac{N}{G} \rceil$ $B_{tile}$.
The key idea of our \benchmark{TW} pattern is to prune each $B_{tile}$ with the regular column pruning (\benchmark{TW-C}) and row pruning (\benchmark{TW-R}) . 

\paragraph{\benchmark{TW-C}.}
To improve the execution efficiency of the proposed sparsity pattern, we first perform the column pruning and for the weight matrix tile $B_{tile}$, which reduces its N-dimension size (i.e., width).
Our approach prunes a different number of columns in each weight matrix tile for better irregularity. 
We use the example in \Fig{fig:tw} \circled{2} to illustrate its advantages.
With the column pruning, the four tiles are pruned with 4, 3, 2, and 1 columns, respectively.
Then, we re-organize the four tiles with $G + 4$, $G + 3$, $G+2$, and $G+1$ columns.
After pruning, the N-dimension sizes of the four tiles are $G$.

\paragraph{\benchmark{TW-R}.}
The row pruning treats an entire row of each weight tile $B_{tile}$ as the basic pruning unit, which leads to the reduced K-dimension size (i.e., height) of each $B_{tile}$.
We prune each $B_{tile}$ with the different number of rows determined by the pruning algorithm that we describe later. 
The difference across different tiles maintains the irregularity of sparsity required by model accuracy. 
E.g., the heights of four weight matrix tiles in \Fig{fig:tw} \circled{2} are $K-2$, $K-4$, $K-8$, and $K-1$ respectively after the row pruning.
The combined row and column pruning alleviates the constraint on the sparsity pattern and therefore allows more weight elements to be pruned.

\paragraph{\benchmark{TEW}.}
Since the \benchmark{TW} still enforces a particular pruning pattern, important weight elements could be removed, leading to accuracy loss.
We propose to overlay \benchmark{TW} and \benchmark{EW} to mitigate the accuracy loss.
\Fig{fig:tw} \circled{3} illustrates the resulted hybrid pattern \benchmark{tile-element-wise (TEW)}. 
In order to prune $\alpha$ percent of weights, the \benchmark{TEW} first prunes $\alpha+\delta$ percent of weights with only \benchmark{TW}, and then restores $\delta$ percent of the weight elements with the highest importance scores. 
For the hybrid \benchmark{TEW} pattern, each tile stores the \benchmark{EW} pattern with the compressed sparse column (CSC) format.
We leverage the linear property of matrix multiplication to execute the \benchmark{TW} and \benchmark{EW} separately.
We explain the execution details in \Sec{sec:implementation}.

\paragraph{\benchmark{TVW}.}
As shown in \Fig{fig:tw} \circled{2}, \benchmark{TW} prunes the rows and columns at the global memory level. The newest Ampere GPU architecture adopts the sparse tensor core with the 2:4 \benchmark{VW} sparsity pattern at the fine-grained register level to prune the two elements of each four-element vector, as illustrated in \Fig{fig:tw} \circled{4}. 

Therefore, the \benchmark{VW} sparsity is orthogonal with the \benchmark{TW} sparsity, and we propose the hybrid \benchmark{TVW} to fuse the \benchmark{VW} and \benchmark{TW} to complement each other. 
First, \benchmark{TW} is a coarse-grained pattern. 
For example, \benchmark{TW-R} prunes a row with $G$ elements, where $G$ is usually greater than $32$. Fusing with \benchmark{VW} can provide a more fine-grained pattern (2-out-of-4) to achieve a more irregular sparsity pattern. 
Second, due to the specificity of hardware design, the sparse tensor core has a fixed $50\%$ sparsity for every four-element vector. \benchmark{TW} has a more uneven sparsity distribution, which can benefit the accuracy, as we explain in ~\Sec{subsec:acc_perf}. Finally, \benchmark{TVW} also has efficient execution supported by the sparse tensor core.

Owing to the existence of globally uneven sparsity distribution and the irregularity inside the vector, \benchmark{TVW} leads to a pattern that is closer to \benchmark{EW} than the \benchmark{VW} pattern. 
\benchmark{TVW}/\benchmark{TW} also removes more weights than \benchmark{BW} owing to its fewer constraints on the pruning shape.

\begin{algorithm2e}[t]
    \caption{The multi-stage pruning algorithm.}
    \label{alg:pruning}
    \small
\KwIn{Trained weight matrix, $W$ with shape ($K$, $N$) \\
\hspace*{1cm}Variable granularity, $G$; 
Target sparsity, $S$; \\
\hspace*{1cm}Sparsity step, $s_s$; \\
\hspace*{1cm}Pattern Pruning Function, PatternPrune.

}

\KwOut{Pruned weight matrix, $w$}
    $w=W$ \\
    $s_t=0$ \\
\While{$s_t<S$}{
    $s_t=$ $s_t + s_s$ \tcp*[h]{Increase the sparsity.} \\
    $w =$ PrunePattern($w, s_t, \delta, K, N, G$) \\
    $w =$ FineTune($w$)\\
}
return $w$
\end{algorithm2e}

\begin{algorithm2e}[t]
    \caption{EW, VW and BW pruning algorithm.}
    \label{alg:ew}
    \small
\KwIn{Weight matrix, $w$, and target sparsity, $s_t$ \\
{\hspace*{1cm}Variable granularity, $G$}
}
\KwOut{Pruned weight matrix, $w$}

\SetKwFunction{EW}{EW}
\SetKwFunction{EWr}{EW-Remedy}
\SetKwFunction{pruning}{Pruning}
\SetKwFunction{VW}{VW}
\SetKwFunction{TW}{TW}
\SetKwFunction{BW}{BW}
\SetKwFunction{TWC}{TW-C}
\SetKwFunction{TWR}{TW-R}
\SetKwFunction{TEW}{TEW}
\SetKwFunction{TVW}{TVW}
\SetKwProg{Fn}{def}{:}{}
\Fn{\EW{$w, s_t$}}{
    $scores=$ ImportanceScoreElement($w$)\\
    $threshold=$ Percentile($scores$, $s_t$)\\
    \While{$e_i \in m$}{
        \If{$scores[i] < threshold$}{
            Prune the element $e_i$\\
        }
    }
    return $w$
}
\Fn{\VW{$w, s_t, {G}$}}{
    
    $w =$ Split $w$ by shape ({$G$}, 1) for VW pruning\\
    $scores=$ {ImportanceScoreVector}($w$)\\
    \While{$v_i \in m$}{
        $threshold_{v_i}=$ Percentile($scores_{v_i}$, $s_t$)\\
        \While{$e_j \in v_i$}{
            \If{$scores_{v_i}[j] < threshold_{v_i}$}{
                Prune the element $e_j$ in $v_i$\\
            }
        }
    }
    return $w$
}
\Fn{\BW{$w, s_t, G$}}{
    
    $w =$ Split $w$ by shape ({$G$}, {$G$}) for BW pruning\\
    $scores=$ ImportanceScoreBlock($w$)\\
    $threshold=$ Percentile($scores$, $s_t$)\\
    \While{$b_i \in m$}{
        \If{$scores[i] < threshold$}{
            Prune the block $b_i$\\
        }
    }
    return $w$
}
\end{algorithm2e}

\begin{algorithm2e}[t]
    \caption{TW-based pruning algorithm.}
    \label{alg:tw}
    \small
\KwIn{Weight matrix, $w$ with shape ($K$, $N$) \\
\hspace*{1cm}Variable granularity, $G$ \\
\hspace*{1cm}Target sparsity, $s_t$
}
\KwOut{Pruned weight matrix, $w$}

\SetKwFunction{EW}{EW}
\SetKwFunction{pruning}{Pruning}
\SetKwFunction{VW}{VW}
\SetKwFunction{TW}{TW}
\SetKwFunction{TWC}{TW-C}
\SetKwFunction{TWR}{TW-R}
\SetKwFunction{TEW}{TEW}
\SetKwFunction{TVW}{TVW}
\SetKwProg{Fn}{def}{:}{}

\Fn{\TW{$w, s_t, K, N, G$}}{
    $s = 1 - \sqrt{(1 - s_t)}$ \\
    $w =$ Split $w$ by shape ($K$, 1) for TW-C\\
    $scores=$ ImportanceScoreVector($w$)\\
    $threshold=$ Percentile($scores$, $s$)\\
    \While{$v_i \in m$}{
        \If{$scores[i] < threshold$}{
            Prune the vector $v_i$ with shape ({$K$}, 1)
        }
    }
    {$w$} = Condense($w$)\\
    $w =$ Split $w$ by shape (1, $G$) for TW-R\\
    $scores=$ ImportanceScoreVector($w$)\\
    $threshold=$ Percentile($scores$, $s$)\\
    \While{$v_i \in m$}{
        \If{$scores[i] < threshold$}{
            Prune the vector $v_i$ with shape (1, {$G$})
        }
    }
    m = Condense($w$)\\
    return $w$
}

\Fn{\TEW{$w, s_t, \delta, K, N, G$}}{
    $s = s_t + \delta$ \\
    $w_s =$ copy($w$) \\
    $w=$ \TW{$w, s, K, N, G$}\\
    $scores=$ ImportanceScoreElement($w_s$)\\
    \While{$e_i \in m$}{
        $scores[index(e_i)] = {0}$
    }    
        $threshold=$ Percentile($scores$, {1 - $\delta$})\\
        \While{$e_i \in m_s $ and $ e_i \notin m$}{
            \If{$scores[i] > threshold$}{
                Remedy the element $e_i$\\
            }
        }    
    return $w$
}
\Fn{\TVW{$w, s_t, K, N, G$}}{
    $s = 1-2*(1-s_t)$ \\
    $w=$ \TW{$w, s, K, N, G$} \\
    \tcp*[h]{VW with fixed 50\% (2:4) sparsity.}\\
    {$w=$ \VW{$w, 0.5, {4}$}} \\
    return $w$
}
\end{algorithm2e}

\section{Tile Sparsity Based Pruning} 
\label{sec:pruning_algorithm}

This section explains our multi-stage pruning algorithm for leveraging the proposed \benchmark{TW} sparsity pattern. 
Algorithm~\ref{alg:pruning} describes the algorithm, which we explain in detail as follows.

\paragraph{Overview.} 
We adopt the multi-stage pruning algorithm that gradually prunes the pre-trained dense model to reach a target sparsity.
Each iteration from Line 4 to 6 in Algorithm~\ref{alg:pruning} is a complete pruning-tuning stage.
Each stage consists of a pruning and fine-tuning step, where the algorithm first prunes the model with a small sparsity target and then fine-tunes the pruned model to restore the model accuracy.
Line 4 increments the sparsity target in the current stage with the sparsity step ($s_s$).
The ``PrunePattern'' in Line 5 corresponds to the functions in Algorithm~\ref{alg:ew} and \ref{alg:tw}, depending on the pruning strategy.
Prior work points out that multi-stage pruning improves the model accuracy more than single-stage pruning~\cite{han2015learning}.

\paragraph{Pattern Pruning.}
Algorithm~\ref{alg:ew} shows the \benchmark{EW}, \benchmark{VW}, and \benchmark{BW} sparsity pattern pruning algorithm.
For all patterns, we first split the weight tensor according to their pruning granularity. For \benchmark{EW}, we regard each element as the individual pruning granularity. And \benchmark{VW} splits by the vector with shape of the $(4,1)$ and \benchmark{BW} with the $(G, G)$ block. Then, we derive their importance score, which is introduced in next.
We can easily sort them and get the pruning threshold by the percentile function with the target sparsity $s_t$. Finally, we can prune the elements (or blocks) according to the threshold. 
Algorithm~\ref{alg:tw} presents the \benchmark{TW}-based algorithms, including \benchmark{TEW} and \benchmark{TVW}. 
As explained in \Sec{sec:tile_sparsity}, the \proj{} pattern requires the column pruning before the row pruning.
We first equally divide the sparsity for the \benchmark{TW-C} and \benchmark{TW-R} (Line 2). Then, we break the weight matrix into column-based tiles (Line 3) and evaluate the importance score of each tile (Line 4).
We then determine the threshold for \benchmark{TW-C} pruning based on the sparsity target caculated before (Line 5).
Line 6-8 remove the column tiles.
Afterward, we reorganize the column-pruned matrix into tiles (line 9).
Line 10-16 performs \benchmark{TW-R} pruning, which is similar to \benchmark{TW-C} pruning.
After the column and row pruning, the weight matrix becomes compatible with \proj{}.
In our algorithm, we design the tiling granularity $G$ as a tunable hyper-parameter, through which we explore the trade-off between the accuracy and performance of the sparse model.
For the \benchmark{TEW} sparsity pattern, we first add more $\delta$ sparsity for \benchmark{TW} (Line 19) and execute the \benchmark{TW} pruning (Line 21). Then we exploit the remedy algorithm (Line 23-28) to restore the $\delta$ \benchmark{EW} sparsity. 
For the \benchmark{TVW} sparsity pattern, we can combine and orderly execute \benchmark{TW} (Line 32) and \benchmark{VW} (Line33) pruning with precise sparsity division (Line 31).

\paragraph{Global Weight Pruning.}
There exists an uneven distribution of weight sparsity in different layers of a DNN model, which previous work~\cite{guo2020accelerating} uses a global weight pruning to exploit.
Taking Algorithm~\ref{alg:tw} as an example, the codes in Line 5 and line 12  sort the scores for all tiles in the column and row pruning, respectively.
The codes in Line 6-9 and Line 13-16 prune the tiles from all layers in the DNN model according to their importance rank.

\paragraph{Importance Score.}
How to compute the importance score is an active research topic~\cite{han2015learning, molchanov2016pruning, molchanov2019importance}.
The most intuitive approach~\cite{han2015learning} is to use the weight's absolute value.
We use a more accurate approach~\cite{molchanov2019importance} that uses the incurred error by removing a parameter as its importance score.

\begin{figure*}[t]
    \hspace{-2mm}
    \includegraphics[width=1\textwidth]{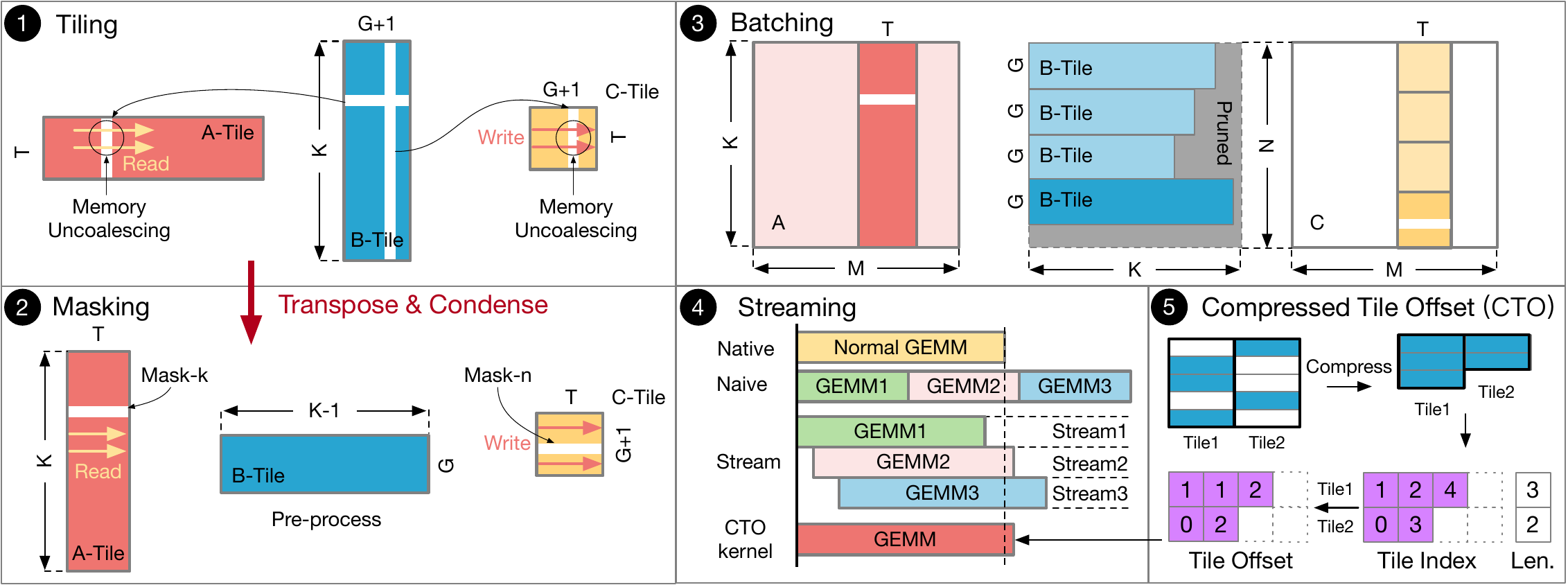}
    \caption{
    Our implementation performs a series of steps to optimize the execution efficiency of \benchmark{TW}-based sparse models. 
    }
    \label{fig:streamgemm}
    \vspace*{-.2cm}
\end{figure*}
    
\section{Efficient GPU Implementation} \label{sec:implementation}

This section introduces our efficient GPU implementation that unleashes the algorithmic benefits of \texttt{TW}. Exploiting the unique sparsity pattern of \texttt{TW}, we first describe the basic tiling design, followed by three key optimizations that combine intelligent data layout and concurrency/batching optimizations to maximize the efficiency of \texttt{TW} tiling on tensor cores.

The advantage of \proj{} sparsity pattern is that sparse matrix multiplication could be transformed to dense GEMM, which can be effectively accelerated on dense GEMM accelerators such as the tensor core on GPUs (\Sec{sec:tile_sparsity}).
\Fig{fig:streamgemm} shows how we transform sparse matrix multiplication that has the \texttt{TW} sparsity pattern to a dense GEMM, and how it exploits various GPU characteristics to maximize the performance.

\paragraph{Tiling.} We start by tiling matrices as usual. \Fig{fig:streamgemm} \circled{1} illustrates an example, where generating an output tile  $C_{tile}$ requires two input tiles $A_{tile}$ and $B_{tile}$. Each input matrix tile has two mask vectors that indicate which rows and columns in the matrix tile are pruned. In the example of \circled{1}, the white rows and columns are pruned. We remove the pruned rows and columns in the weight matrix tile $B_{tile}$, which can be done offline before the model inference starts.
The input tile $A_{tile}$ and output tile $C_{tile}$ are stored in the original dense format. Their pruned rows/columns are skipped rather than removed.

We modify the dense GEMM kernel such that it skips computing partial sums for pruned elements according to the mask vectors. This reduction of computation is the source of acceleration. Our baseline GEMM implementation is based on the open-sourced CUTLASS~\cite{cutlass2019}, which is a high-performance linear algebra CUDA library. It implements three levels of tiling to maximize the data reuse in the global memory (thread block tile), shared memory (warp tile), and register file (thread fragment).
Meanwhile, it can also leverage the tensor core in the GEMM computation, which we use to accelerate the \proj{}.

However, a naive tiling implementation is inefficient and even causes slowdown compared to the original dense model. In our implementation, we exploit three optimizations that mitigate the inefficiencies and maximize the benefits of \texttt{TW}.

\paragraph{Memory Accesses Coalesce.} Naive tiling leads to frequent uncoalesced memory accesses that are inefficient on the single-instruction-multiple-data based GPUs. \Fig{fig:streamgemm} \circled{1} shows the memory access patterns in the original row-major matrix format.
The pruned row in $B_{tile}$ causes the skip of column in $A_{tile}$.
Therefore, a continuous access to the $A_{tile}$ (marked by the yellow arrows) that is originally coalesced now becomes uncoalesced, which can cause severe performance degradation as uncoalesced memory accesses require multiple memory transactions.
The uncoalesced accesses also exist in the matrix tile $C_{tile}$ (marked by the red arrows) owing to the pruned column in $B_{tile}$.

We propose to store the matrix tiles in their transposed format to optimize their memory access efficiency.
In \Fig{fig:streamgemm} \circled{2} where the three tiles are transposed, the column skipping is converted to the row skipping. 
Thus, it eliminates the uncoalesced accesses and improves the access efficiency.

\begin{figure}[t]
    \centering
    \includegraphics[width=1\columnwidth]{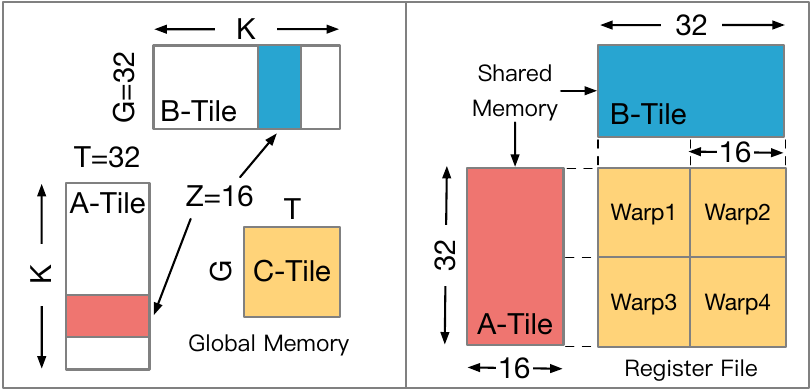}
    \caption{Warp-level GEMM tiling that exploits tensor core.}
    \label{fig:warplevel}
    \vspace*{-.6cm}
\end{figure}

\paragraph{Load Imbalance Mitigation.}  \texttt{TW} sparsity inherently introduces imbalanced tiles. That is, some tiles will require more computations since fewer rows are pruned; other tiles that have more rows pruned will lead to lower computation. Imbalanced tiles lead to resource under-utilization, and thus affects the overall speedup.
We propose to batch tile computations to improve the utilization. \Fig{fig:streamgemm} \circled{3} shows an example where the weight matrix is decomposed into $\lceil \frac{N}{G} \rceil$ tiles, where $G$ is the \texttt{TW} granularity.
Different $B_{tile}$ are batched together to share the same $A_{tile}$. Batching improves resource utilization as a batched GEMM packs multiple tiles and thus increases the computation.

Another practical benefit of batched-GEMM implementation is that we can reuse existing high-performance tensor core-based GEMM kernels and avoid implementing specialized GEMM kernels, each customized for a particular tile size.
\Fig{fig:warplevel} illustrates the warp-level tiling and Listing~\ref{lst:gemm} shows the kernel implementation that uses tensor core APIs.
We assume that $G=32$, $T=32$ and $Z=16$, which is the minimum tiling granularity as it must be the multiple of $32$ (i.e., warp size). $A_{tile}$ and $B_{tile}$ are stored into the shared memory and $C_{tile}$ to the register file. Then a warp tile will compute the out-product with the tensor core MMA API, which can support the fixed size ($16 \times 16\times 16$) matrix multiplication.

{
\begin{lstlisting}[float=t, language=C++, caption={CTO-based GEMM kernel on tensor core.}, label={lst:gemm}]
#define G 32
#define T 32
#define Z 16
__global__ void CTOGEMM(int M, int Pruned_N, int Pruned_K, half *A, half *B, half *C, half alpha, half beta, int *CTO_n, int *CTO_k){    
    //Allocate C_tile in Register File.
    half C_tile[G * T];
    //Allocate A_tile and B_tile in Shared Memory.
    __shared__ half A_tile[T * Z];
    __shared__ half B_tile[G * Z];
    for(int k = 0; k < K; k += Z){
        //Load A_tile from Global to Shared Memory skipping the pruned row with CTO_k.
        Load_A_Tile_with_Mask(A_tile, A, CTO_k);  
        //Load B_tile from Global to Shared Memory with Pre-Processed B.
        Load_B_Tile(B_tile, B);
        //Tensor core API: WMMA with fixed 16x16x16 GEMM.
        WMMA::MMA(C_tile, A_tile, B_tile, alpha, beta);
    }
    //Store C_tile from Register File to Global Memory skipping the pruned row with CTO_n.
    Store_C_Tile_with_Mask(C, C_tile, CTO_n);
}
\end{lstlisting}
}

While batching mitigates resource under-utilization, we find that it is possible that the computation of a batch still under-utilizes the GPU resources. Our previous work~\cite{guo2020accelerating} leverages concurrent kernel execution on modern GPUs~\cite{streams2015} to further improve resource utilization. 
In the studied NVIDIA GPU platform, we overlap the computation of different tiles by assigning to different streams, and rely on the underlying scheduler to maximize resource utilization. \Fig{fig:streamgemm} \circled{4} shows an example where naively running different batches could have lower performance than the original unpruned GEMM. Concurrently executing multiple batches with different streams improves performance.

\paragraph{{Tile Fusion and Compressed Tile Offset.}}
The stream GEMM  in \Fig{fig:streamgemm} \circled{4} launches multiple kernels, each with two mask vectors. And all kernels execute concurrently within different streams. 
The resource utilization can be further improved by fusing the computation of all tiles into only one kernel. 
Inspired by the compressed sparse column (CSC) method, as shown in \Fig{fig:streamgemm} \circled{5}, we change the tile mask into the index of the unpruned row/column. As such, the tile index will have more memory efficiency compared to the tile mask when the sparsity increases.

First, we fuse all tile computations with a two-dimension tile index with tile length number, as shown in \Fig{fig:streamgemm} \circled{5}. For example, the two tiles have three rows (1, 2, 4) and two rows (0, 3) to compute, respectively.
Instead of using multiple tile masks to indicate which columns are pruned for each tile, we merge and pad them into a matrix and launch one kernel for the GEMM. 
This method takes full advantage of the concurrency and allows schedulers to maximize resource utilization at the thread block level.

Second, we continue to change the indices of row/column to the offsets, which are friendly to the GPU global memory access. For example, the GPU needs to access the first tile with original indices (0, 1, 2). We only add the offsets (1, 1, 2) with the original indices and change them to (1, 2, 4). Then the GPU can efficiently and correctly find the memory address. 
The tile offsets are friendly to the global memory access mechanism and improve memory access efficiency.


\section{Evaluation} \label{sec:evaluation}

In this section, we demonstrate that \proj{} is able to maintain the accuracy of sparse DNN models and provide the significant execution speedup over the dense model and other sparsity patterns at the same time. 
We first explain our evaluation methodology with the use of state-of-the-art DNN models on the GPU equipped with (sparse) tensor cores.
We then study the design space of \proj{} to explore the trade-off between model accuracy and latency.
In the end, we select the representative configurations of \proj{} and \benchmark{TVW} and compare it with other sparsity patterns.

\subsection{Methodology}

\paragraph{Benchmark.} We evaluate five popular neural networks, VGG16, ResNet-18, ResNet-50 (CNN), NMT (LSTM), and BERT (Transformer), which cover tasks from the computer vision and NLP domain.VGG16~\cite{simonyan2015very}, ResNet-18~\cite{he2016deep}, and ResNet-50~\cite{he2016deep} are popular CNN models.
We evaluate its accuracy for image classification on the ImageNet~\cite{krizhevsky2017imagenet} dataset with 1.2 million training images and 50,000 validation images.
We prune its weight matrix after applying the \benchmark{img2col} method~\cite{chetlur2014cudnn}, which flattens the filters in the same channel to a column and different columns correspond to different channels (so the left flattened feature map matrix multiplies the flattened weight matrix in \Fig{fig:tw}).
This approach is similar to prior work\mbox{~\cite{zhu2019sparse}}.

We evaluate the accuracy of NMT model, which adopts the attention based encoder-decoder architecture, for the machine translation task.
We reproduce the model with an open-source framework~\cite{luong17}.
We evaluate the NMT model on the IWSLT English-Vietnamese dataset~\cite{luong2016acl_hybrid}, and use the BLEU (bilingual evaluation understudy) score~\cite{papineni2002bleu} as the accuracy metric.
For the Transformer model family, we use the BERT-base with $12$ layers.
The two evaluated downstream tasks are the sentence-level classification on the widely used GLUE (general language understanding evaluation) dataset~\cite{wang2019glue} and the more challenging question answering task on the SQuAD dataset~\cite{rajpurkar2016squad}.
The GLUE dataset is a composite dataset with 10 different sub-tasks, and we evaluate 6 out of them.

In our experiments, we use the pre-trained models that can achieve their reported reference accuracies.
We then apply \benchmark{EW}, \benchmark{VW}, \benchmark{BW}, and our proposed \benchmark{TW}, \benchmark{TEW}, and \benchmark{TVW} sparsity patterns to prune the dense models according to the algorithm described in \Sec{sec:pruning_algorithm}. 
To conduct a fair comparison, we strictly set the same hyper-parameters (such as learning rate) for all patterns and fine-tune with the same epochs for each sparsity step.
We use the PyTorch~\cite{paszke2019pytorch} and TensorFlow~\cite{tensorflow2015-whitepaper} framework for fine-tuning. Depending on the dataset size, we perform the fine-tuning for 4-10 epochs at each target sparsity level for BERT and NMT, which is sufficient to saturate the model accuracy in our experiment. For CNNs, we follow the pruning work~\cite{lin20221xn} to fine-tune the pruned models with 100 epochs.

\begin{figure*}[t]
    \begin{subfigure}{0.33\textwidth}
        \includegraphics[width=\linewidth]{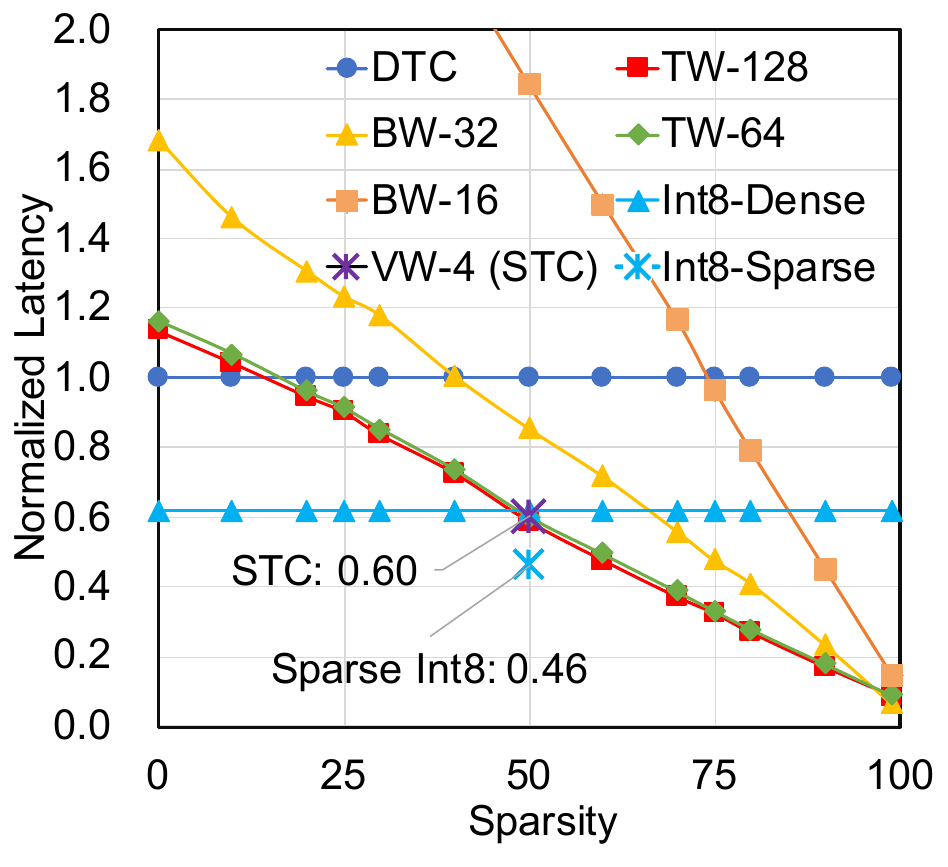}
        \caption{{\hlc{Normalized latency on tensor core.}}}
        \label{fig:g_latency_t}
    \end{subfigure}
    \begin{subfigure}{0.33\textwidth}
        \includegraphics[width=\linewidth]{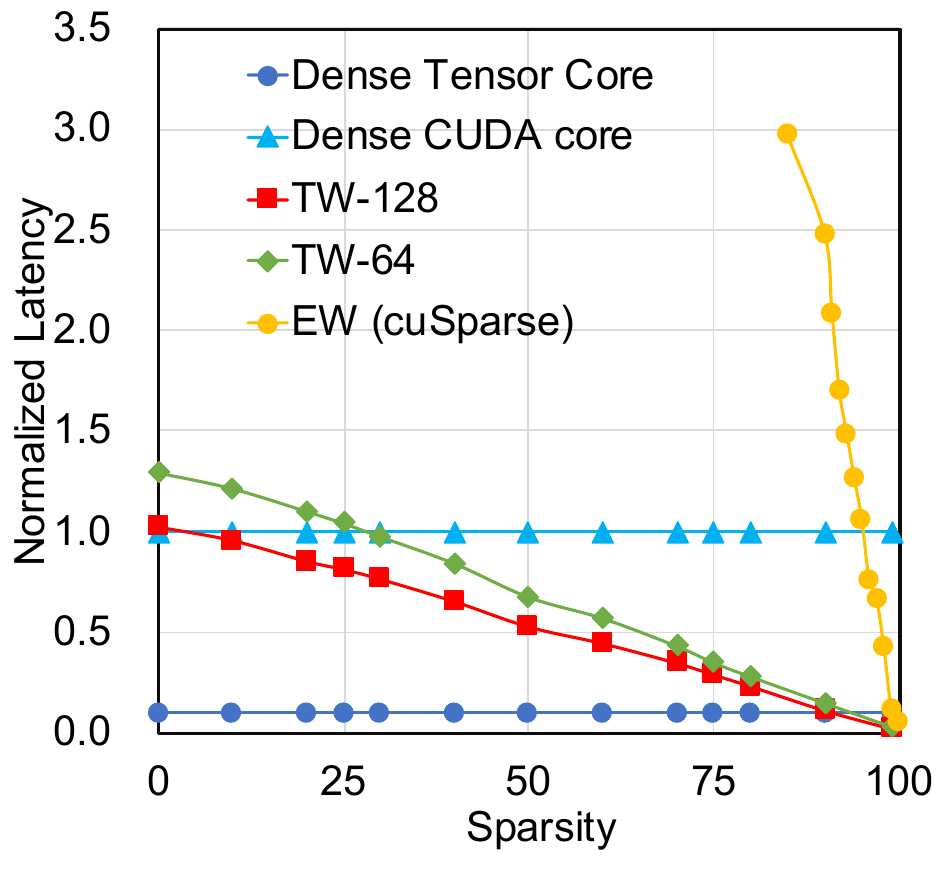}
        \caption{{\hlc{Normalized latency on CUDA core.}}}
        \label{fig:g_latency_c}
    \end{subfigure}
    \begin{subfigure}{0.33\textwidth}
        \includegraphics[width=\linewidth]{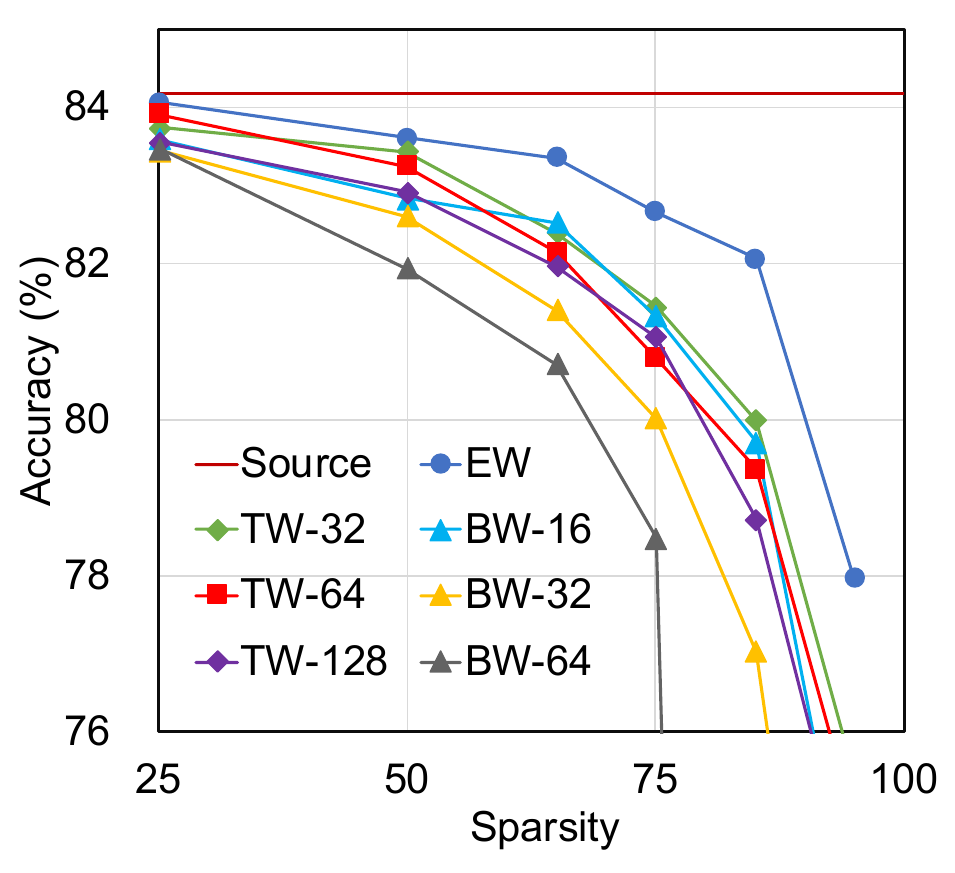}
        \caption{Accuracy of BERT on MNLI.}
        \label{fig:g_accuracy}
    \end{subfigure}~
	\caption{ 
        Normalized latency comparison of ($4096 \times 4096 \times 4096$) GEMM. DTC and STC represent dense and sparse tensor core, respectively.
         Figure (c) compares the accuracy under different pruning granularities. 
    }
	\label{fig:g_impact}
    \vspace*{-.6cm}
\end{figure*}

\paragraph{Baselines.} 
We compare the proposed \benchmark{TW}, \benchmark{TEW}, and \benchmark{TVW} with \benchmark{EW}, \benchmark{VW}, and \benchmark{BW}.
For accuracy, we evaluate all patterns on the DNN models after fine-tuning. Especially for the \benchmark{VW}, there are two different settings. First, the sparse tensor core of the latest NVIDIA Ampere GPU architecture~\cite{a100} adopts the 2:4 (2-out-of-4) pattern with fixed 50\% sparsity. Second, the previous research~\cite{zhu2019sparse} proposed another type of sparse tensor core~\cite{zhu2019sparse}, which has the 4:16 (4-out-of-16) pattern with a fixed 75\% sparsity. As such, we conduct two settings for \benchmark{VW}, \benchmark{VW-4} for the real GPU A100 with 2:4 sparsity and \benchmark{VW-16} for n:16 sparsity, which we only use for accuracy comparison because it can not be accelerated supported by the existing GPUs.

For the latency evaluation, we execute  \benchmark{EW} and \benchmark{BW} using the latest cuSparse~\cite{nvidia2019toolkit} library. We also implement \benchmark{BW} with the BlockSparse library in the Triton~\cite{tillet2019triton}. Surprisingly, the two kinds of block sparsity implementation (i.e., Triton and cuSparse) perform similarly because they use the same algorithm and programming model based on CUTLASS~\cite{cutlass2019}. Finally, we choose the cuSparse as our baseline in this paper.

Our \benchmark{TW}-based implementation (\Sec{sec:implementation}) is based on CUTLASS~\cite{cutlass2019}, an open-source, high-performance GEMM template library. 
For the \benchmark{VW-4}, we evaluate it on the GPU A100 with CUTLASS sparse implementation for the sparse tensor core. For \benchmark{TVW}, we combine and implement the \benchmark{TW} patterns on the \benchmark{VW}-based sparse version of CUTLASS. 
For all those libraries, including \proj{}, we modify the original model codes to call each library explicitly. In this paper, we focus on the GEMM execution time.

We conducted on the Tesla A100 GPU~\cite{a100}, which is added with the sparse tensor core. Therefore, we only evaluate the \benchmark{VW-4} benchmark on A100.
The \benchmark{EW} runs only on the CUDA core with the cuSparse library and the \benchmark{BW} implementation can run on the tensor core supported by cuSparse library.
The convolution operations in the CNN workloads are converted to GEMM by the \benchmark{img2col} method~\cite{chetlur2014cudnn}. The models are all trained using FP32. All inferences on the CUDA core are done using FP32, and all inferences on the tensor core are done using FP16.

We label all pruning patterns by \benchmark{XX-YY}, where \benchmark{XX} represents the pruning pattern, and \benchmark{YY} is the granularity.
For example, \benchmark{TW-64} means \benchmark{TW} adopts granularity $G=64$.
In particular, \benchmark{VW-4} and \benchmark{VW-16} represent \benchmark{VW} with the 2:4 pattern and n:16 pattern, respectively.


\subsection{Design Space Exploration}

We now study the design space of \proj{}, which is the tiling granularity $G$, to explore the trade-off between model accuracy and latency.
In addition, we also evaluate the hybrid \benchmark{TEW} pattern, which extends the trade-off space in sparse models.

We first explore the impact of tiling granularity $G$ for \benchmark{TW} and \benchmark{BW}  pruning.
As shown in \Fig{fig:g_latency_t} and \Fig{fig:g_latency_c}, we compare the normalized latency of \benchmark{EW}, \benchmark{VW}, \benchmark{BW}, and \benchmark{TW}. \benchmark{VW} and \benchmark{BW} can only be supported by the (sparse) tensor core of A100. \benchmark{EW} can only run on the CUDA core implemented by cuSparse. Our proposed \benchmark{TW} and \benchmark{TVW} pattern can run on both tensor cores and CUDA cores. All experiments are conducted on the GEMM with shape ($4096 \times 4096 \times 4096$).

\paragraph{Speedup.} 
The tiled GEMM performance greatly corresponds to the tiling size (granularity). Smaller tiling size leads to less computation in the thread-block (SM) level tile but more global memory accesses, leading to a lower utilization ratio of SM and degraded performance of GEMM, and vice versa.
\benchmark{TW-128} and \benchmark{TW-64} have similar latency-sparsity trends because we optimize the tiling size settings. As illustrated in \Fig{fig:warplevel}, we can adjust the  $T$ of $A_{tile}$ corresponding to the granularity $G$ of $B_{tile}$. $T$ and $G$ are independent due to the design of \benchmark{TW}. $T$ in \benchmark{TW-64} is twice as big as that in \benchmark{TW-128}, then they have the same computation operations and global memory usage for each thread-block (SM) level tile.
\benchmark{TW-128} can surpass the performance of dense GEMM when the sparsity threshold is just larger than 10\% and 5\% on the tensor core and CUDA core, respectively.
In contrast, the sparsity threshold of \benchmark{BW-32} and \benchmark{BW-16} are 40\% and 70\% on the tensor core due to their smaller tiling size ($32$ or $16$). 
\benchmark{VW-4} is exactly the GEMM on the sparse tensor core  and achieves the fixed $1.67\times$ speedup compared to the dense GEMM on the tensor core.
\benchmark{EW} needs more than 95\% sparsity to be better than the dense GEMM on the CUDA core.
We also combine the dense tensor core results normalized to the dense CUDA core in the \Fig{fig:g_latency_c}. 
There is a significant advantage (about $9.7\times$ speedup) of the tensor core over the CUDA core because GPU A100~\cite{a100} provides $16\times$ compute capability (TOPs, tera operations per second) over the CUDA core (19.5 TOPs FP32) for the tensor core (312 TOPs FP16).
In summary, with the optimizations in \Sec{sec:implementation}, the  \benchmark{TW} achieves the best performance on the tensor core and CUDA core among all sparsity patterns. 

\paragraph{Int8 Quantization.} 
We also compare the performance against the quantization method: \benchmark{Int8} quantization (dense) and \benchmark{Int8} quantization with \benchmark{VW} sparsity (sparse), i.e., sparse tensor core with \benchmark{Int8} quantization. The \benchmark{Int8-Dense} achieves $1.62\times$ speedup over the dense GEMM and is similar to \benchmark{VW-4} because they have exactly equivalent memory footprint and computation load, which are 50\% of original FP16 models. The \benchmark{Int8} with \benchmark{VW-4} sparsity can achieve a further $2.16\times$ speedup with 25\% memory and computation of FP16 models.

\begin{figure}[t]
    \centering
    \begin{subfigure}{0.23\textwidth}
      \includegraphics[width=\linewidth]{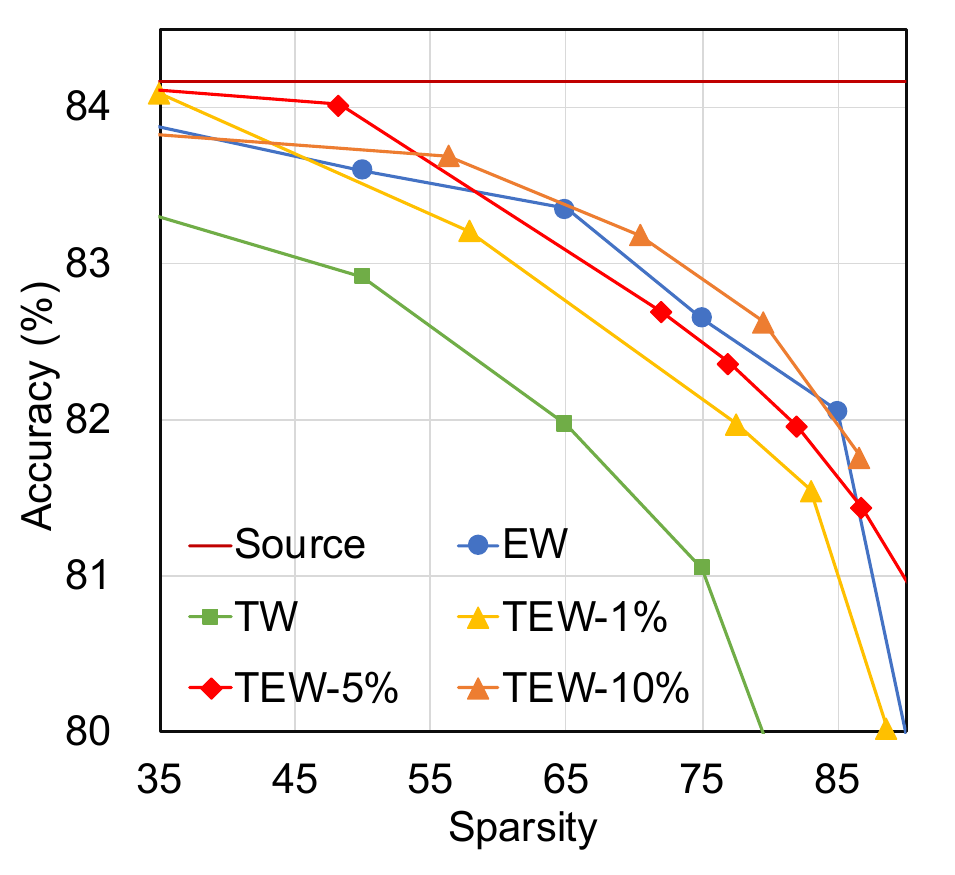}
    \caption{Accuracy.}
    \label{fig:delta_accuracy}
    \end{subfigure}
    \begin{subfigure}{0.25\textwidth}
             \includegraphics[width=\linewidth]{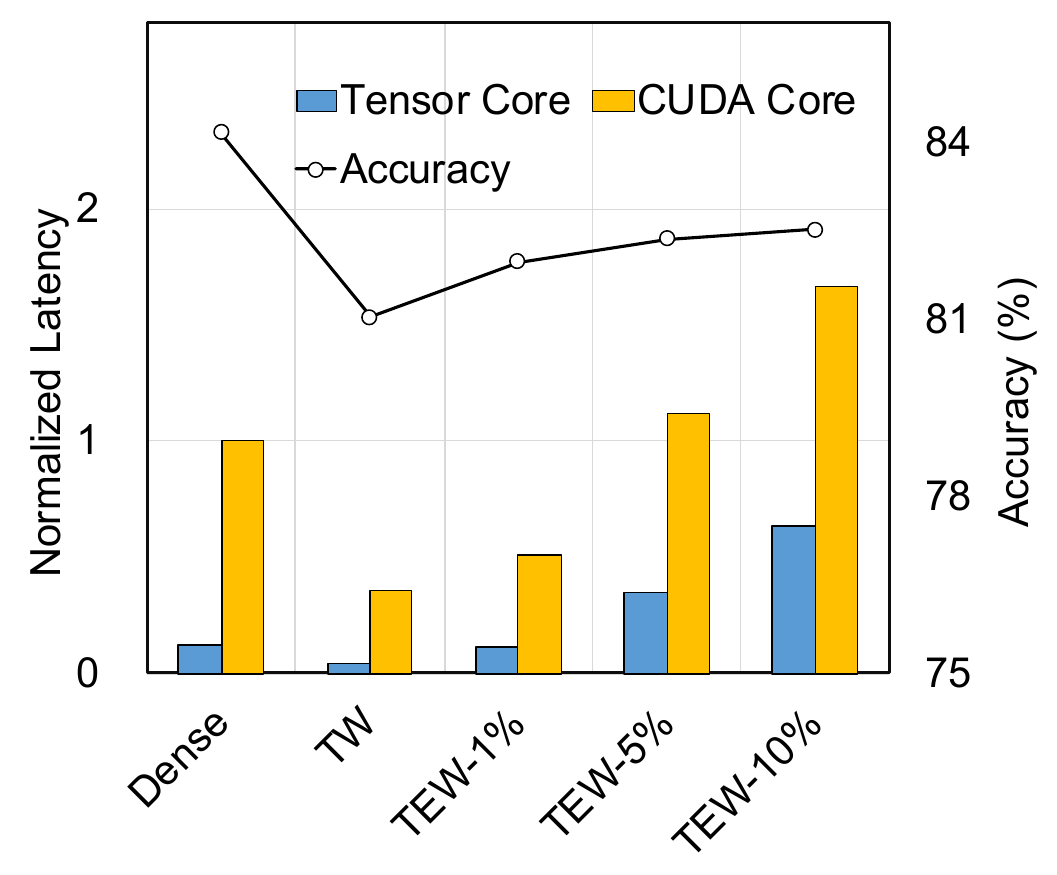}
    \caption{\hlc{Normalized latency.}}
        \label{fig:delta_latency}
    \end{subfigure}
	\caption{Accuracy and latency of \benchmark{TEW}-based sparse BERT model with different $\delta$ values, which determine the portion of added \benchmark{EW} elements. All latency values in (b) are normalized to the latency of dense model on CUDA core.}
	\label{fig:delta_impact}
    \vspace*{-0.5cm}
\end{figure}

\paragraph{Accuracy.} 
\Fig{fig:g_accuracy} compares the accuracy of \benchmark{EW}, \benchmark{BW}, and \benchmark{TW}. 
The analysis is case-studied on the BERT model for sentence pair entailment task on the MNLI dataset.
The most fine-grained \benchmark{EW} achieves the best model accuracy as expected.
When sparsity is less than $50\%$, all the granularities evaluated have similar accuracies except \benchmark{BW-64}, suggesting that the BERT model is at least 50\% redundant. In particular, at a sparsity of 75\%, our proposed \texttt{TW-128} has an accuracy loss of about 1.6\% compared to \texttt{EW}. As the sparsity increases, the accuracy drop becomes more significant. The most coarse-grained \benchmark{BW-64} experiences the most drastic accuracy drop of $>5\%$ at 75\% sparsity. The accuracy of \benchmark{BW-64} is unacceptable for the DNN model, and \benchmark{BW-32} also has a gap compared to \benchmark{TW}. 
The accuracy drop of \texttt{TW} increases slightly with a larger $G$ value. This is because the larger $G$ value puts a more strict constraint on the pruning shape, but larger $G$ also means greater latency reduction.
We find that $G$ of 128 and 64 are sufficient to maintain the model accuracy while providing significant latency reduction.

This comparison shows that \benchmark{TW} has a significant advantage over other sparsity patterns. To reduce the experiment exploration space, we first make \benchmark{TW} and \benchmark{BW} have similar accuracy trends and then compare their latency. Therefore, we set \benchmark{BW} with $G=16$ for the following experiments and set \benchmark{TW}/\benchmark{TVW} with $G=64$ for CNN models and $G=128$ for NMT and BERT models in the rest experiments of this section.


\begin{figure*}[t]


    \begin{subfigure}{0.33\textwidth}
        \includegraphics[trim=0 0 0 0, clip,  width=\linewidth]{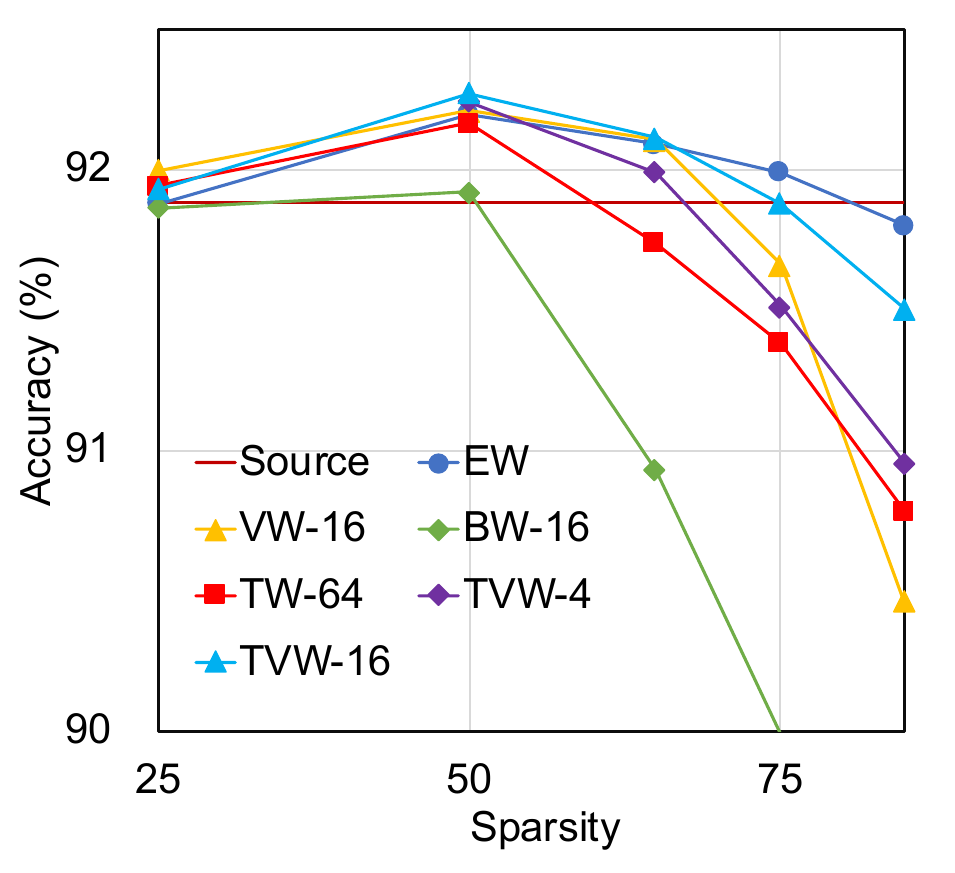}
        \vspace*{-0.7cm}
        \caption{VGG-16}
        \label{fig:vgg_acc}
        \end{subfigure}~
    \begin{subfigure}{0.33\textwidth}
        \includegraphics[trim=0 0 0 0, clip,  width=\linewidth]{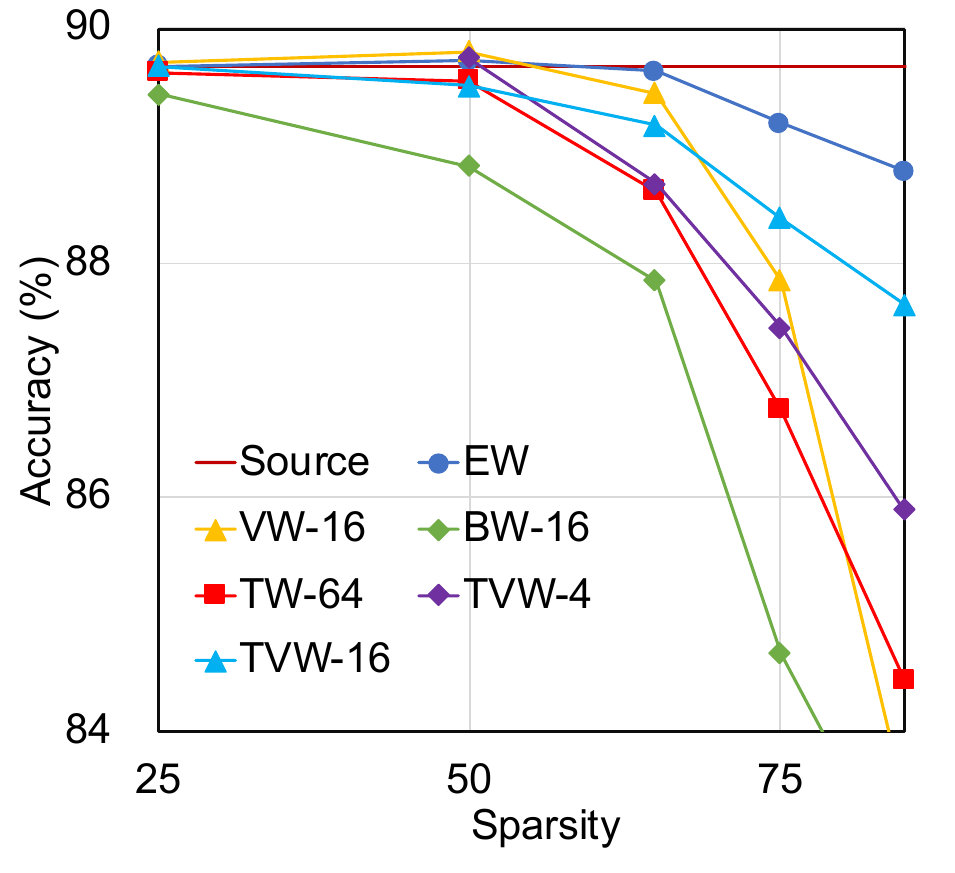}
        \vspace*{-0.7cm}
        \caption{ResNet-18}
        \label{fig:resnet18_acc}
    \end{subfigure}~
    \begin{subfigure}{0.33\textwidth}
        \includegraphics[trim=0 0 0 0, clip,  width=\linewidth]{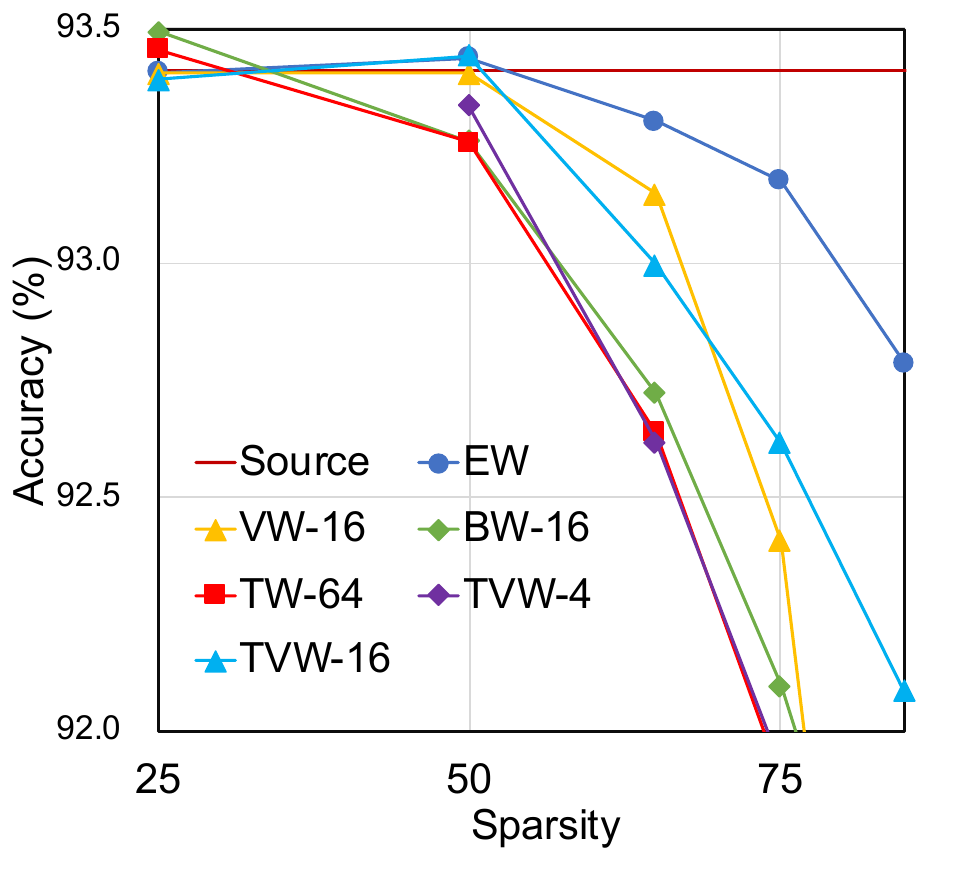}
        \vspace*{-0.7cm}
        \caption{ResNet-50}
        \label{fig:resnet50_acc}
    \end{subfigure}
        \vfill
    \begin{subfigure}{0.33\textwidth}
        \includegraphics[trim=0 0 0 0, clip, width=\linewidth]{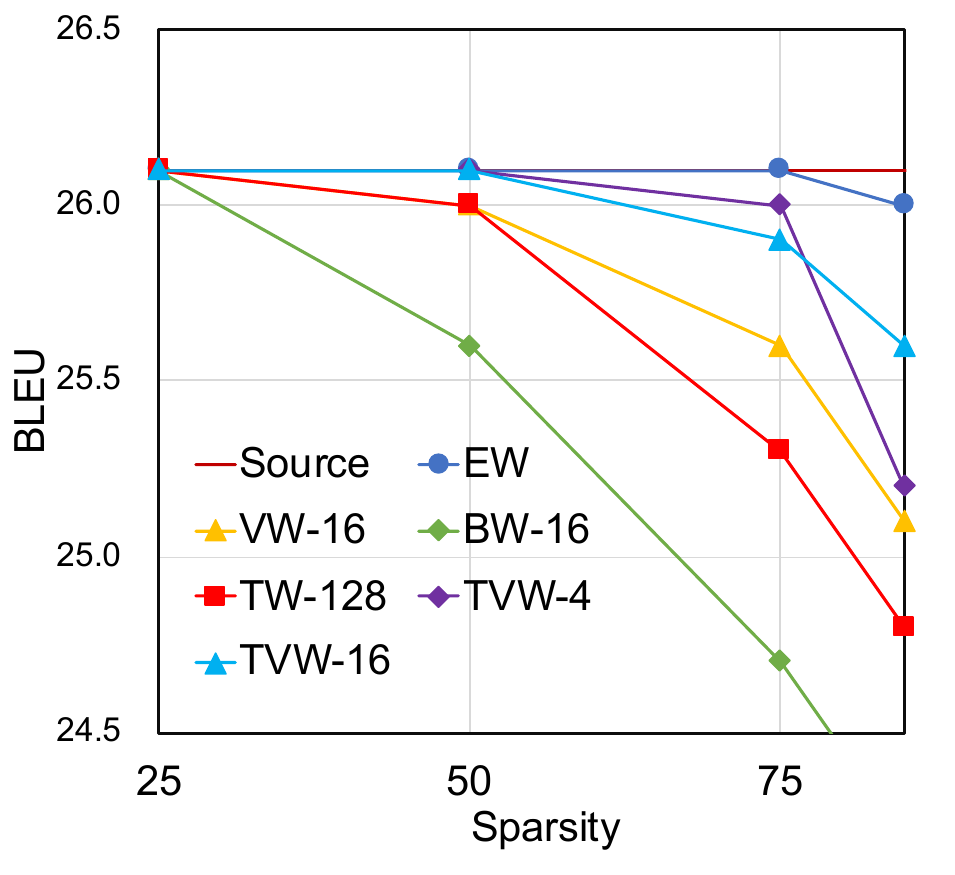}
        \vspace*{-0.7cm}
        \caption{NMT}
        \label{fig:nmt_acc}
        \end{subfigure}~
      \begin{subfigure}{0.33\textwidth}
      \includegraphics[trim=0 0 0 0, clip, width=\linewidth]{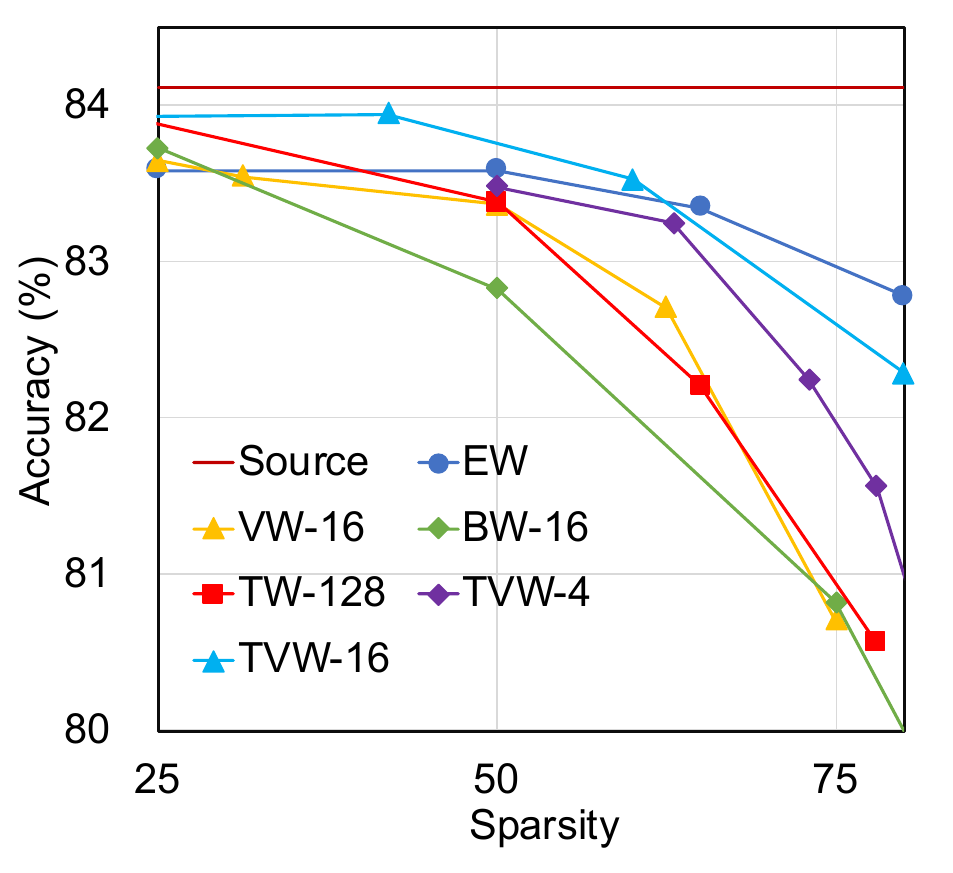}
      \vspace*{-0.7cm}
      \caption{BERT on MNLI}
      \label{fig:mnli_acc}
      \end{subfigure}~
      \begin{subfigure}{0.33\textwidth}
      \includegraphics[trim=0 0 0 0, clip,  width=\linewidth]{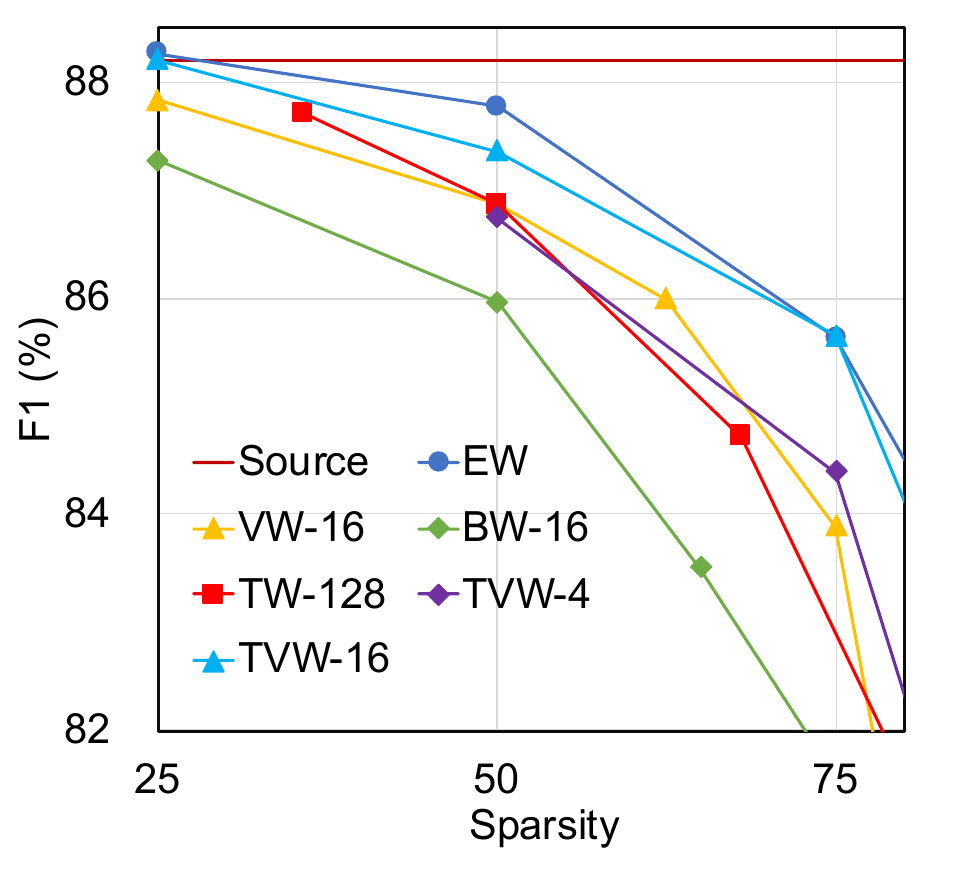}
          \vspace*{-0.7cm}
      \caption{BERT on SQuAD}
      \label{fig:squad_acc}
      \end{subfigure}

      \caption{The accuracy comparison of different models with various pruning patterns and varying sparsity levels.}
      \label{fig:accuracy_all}
     \vspace*{-0.6cm}
  \end{figure*}

\paragraph{Impact of $\delta$ in \benchmark{TEW}.}
We evaluate the impact of $\delta$ in \benchmark{TEW}, which determines the amount of \benchmark{EW} pattern imposed on \texttt{TW} (\Sec{sec:tile_sparsity}).
\mbox{\Fig{fig:delta_accuracy}} compares the sparse BERT model accuracy of different sparsity levels with \mbox{\benchmark{EW}, \benchmark{TW}, and \benchmark{TEW}} patterns.
The accuracy of the sparse model with \mbox{\benchmark{TW}} is lower than \mbox{\benchmark{EW}}.
On the other side, \benchmark{TEW} can mitigate the accuracy loss in \benchmark{TW} by adding a small portion \benchmark{EW}, which is controlled by the $\delta$ parameter in \Sec{subsec:tw}.
For instance, with $\delta = 5\%$, the \benchmark{TEW} accuracy catches up with \benchmark{EW}, and \benchmark{TEW-10\%} surpasses \benchmark{EW}.

\Fig{fig:delta_latency} compares the latency  (left $y$-axis) and accuracy (right $y$-axis) of the dense model and various \benchmark{TW} and \benchmark{TEW} models with the fixed 75\% sparsity. We show the latency results on both tensor core and the CUDA core, which are all normalized to the dense model latency on CUDA core.
On the tensor core, \benchmark{TW} achieves \hlc{$2.98 \times$} speedup than the dense model. \benchmark{TEW} achieves no speedup at $\delta=1\%$ compared to the dense model, and its performance is worse as $\delta$ increases. This is because the irregular portion of \benchmark{TEW} (i.e., the \texttt{EW} portion) could not be executed on the dense tensor cores and, instead, has to be executed on the CUDA cores, which is about 8$\times$ slower than the tensor cores. To illustrate the point, we show the results of running different sparse models on CUDA cores only.
Using CUDA cores alone, \benchmark{TEW} with $\delta=1\%$ is about $2\times$ faster than the dense model.
Thus, we expect that \benchmark{TEW} is useful in resource-constraint scenarios such as mobile systems.

\subsection{Accuracy Comparison}
\label{subsec:acc_perf}
We compare the accuracy and latency speedup of \benchmark{TW} with \benchmark{EW}, \benchmark{VW}, \benchmark{BW} on three different models. 
We perform the comprehensive evaluation of BERT model for the sentence classification task on the composite GLUE dataset, which includes ten different datasets.
We observe similar results on 6 studied datasets and therefore only report the result on the largest dataset MNLI.
We also report its result on the question answering task with the SQuAD dataset.
For the latency speedup, we report the results on the A100 GPU using tensor core and CUDA core separately.

\begin{figure*}[t]
    \vspace*{-0.5cm}
    \hspace{-0.6cm}
       \begin{subfigure}{0.22\textwidth}
       \includegraphics[width=\linewidth]{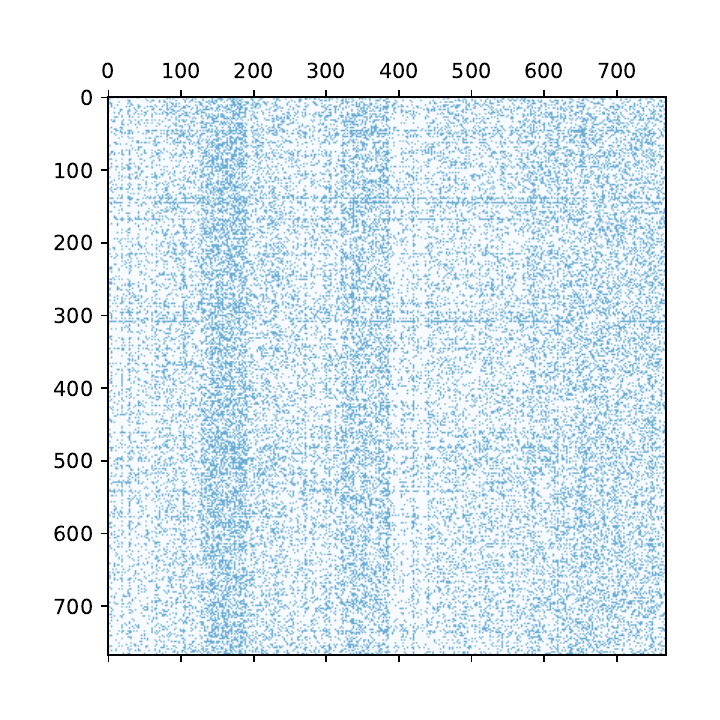} 
       \vspace*{-0.8cm}
       \caption{\benchmark{EW}}
       \label{fig:pattern_a}
       \end{subfigure}~
       \hspace{-0.5cm}
       \begin{subfigure}{0.22\textwidth}
       \includegraphics[width=\linewidth]{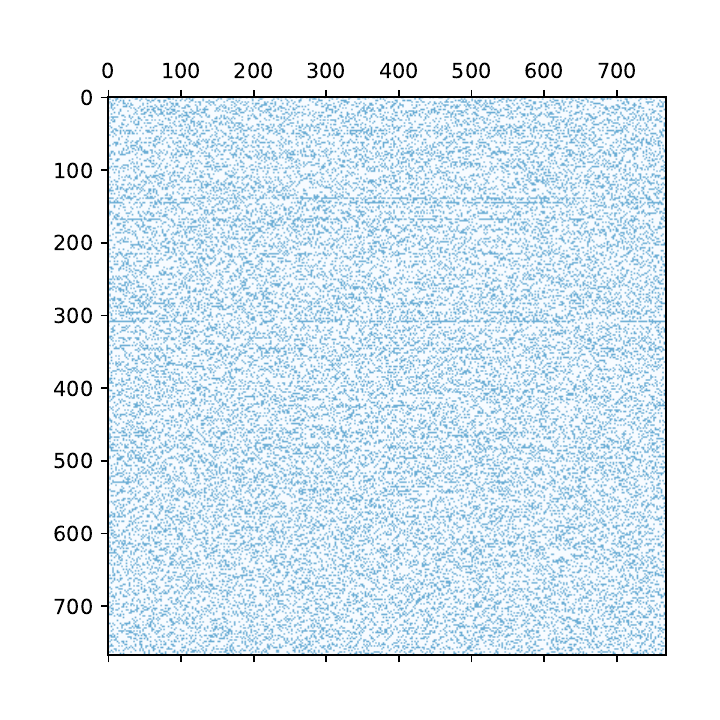}
       \vspace*{-0.8cm}
       \caption{\benchmark{VW-16}}
       \label{fig:pattern_b}
       \end{subfigure}~
       \hspace{-0.5cm}
       \begin{subfigure}{0.22\textwidth}
       \includegraphics[width=\linewidth]{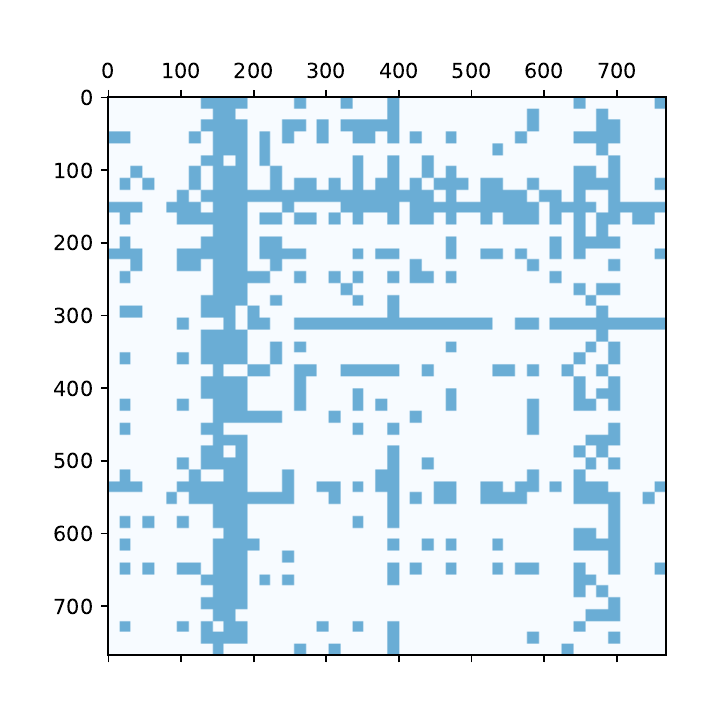} 
       \vspace*{-0.8cm}
       \caption{\benchmark{BW-16}}
       \label{fig:pattern_c}
       \end{subfigure}~
       \hspace{-0.5cm}
       \begin{subfigure}{0.22\textwidth}
       \includegraphics[width=\linewidth]{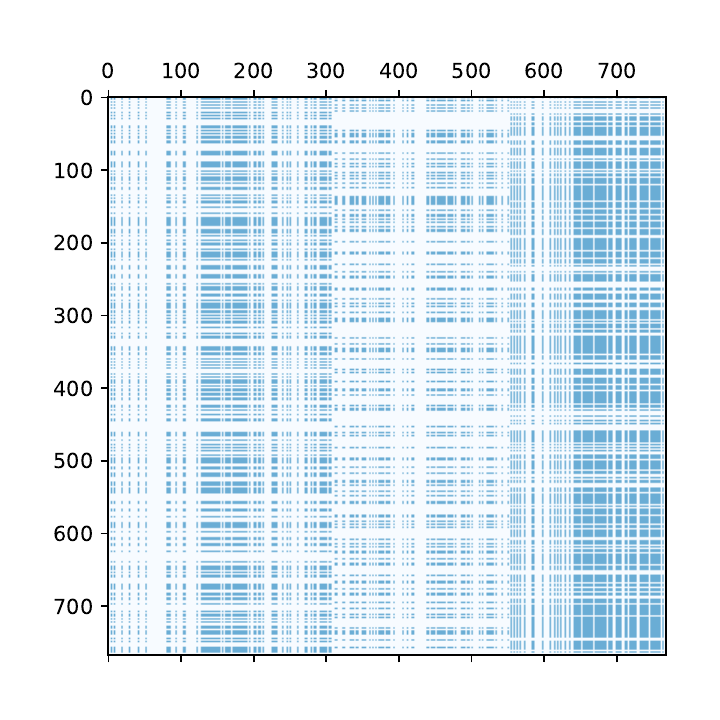}
       \vspace*{-0.8cm}
       \caption{\benchmark{TW-64} }
       \label{fig:pattern_d}
       \end{subfigure}~   
       \hspace{-0.5cm}
       \begin{subfigure}{0.22\textwidth}
       \includegraphics[width=\linewidth]{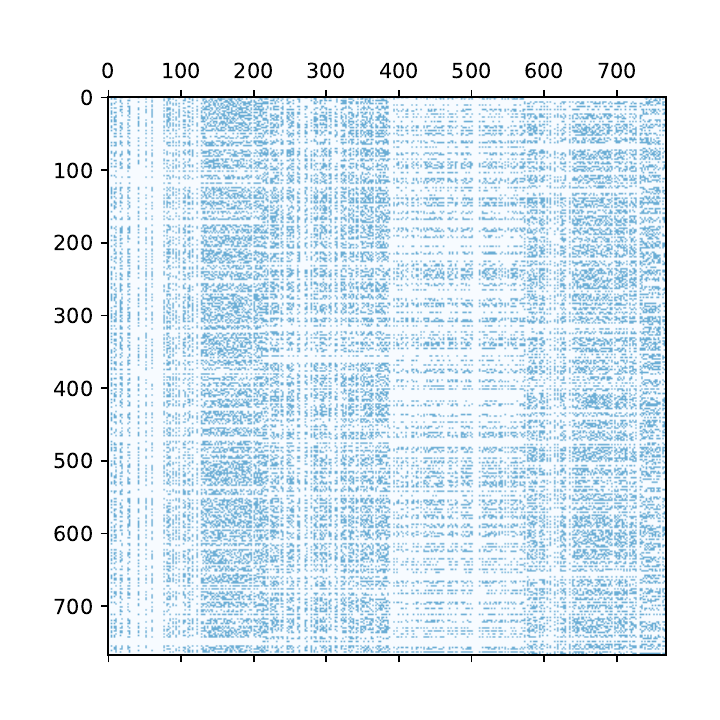}
       \vspace*{-0.8cm}
       \caption{\benchmark{TVW-4}}
       \label{fig:pattern_e}
       \end{subfigure}     
        \vspace*{-0.1cm}
       \caption{Different pruning patterns under 75\% sparsity on first layer self-attention weight matrix $\omega_Q$ in BERT model.}
       \label{fig:pattern}
       \vspace*{-0.6cm}
   \end{figure*}

\begin{figure*}[b]


    \begin{subfigure}{0.33\textwidth}
        \includegraphics[trim=0 0 0 0, clip,  width=\linewidth]{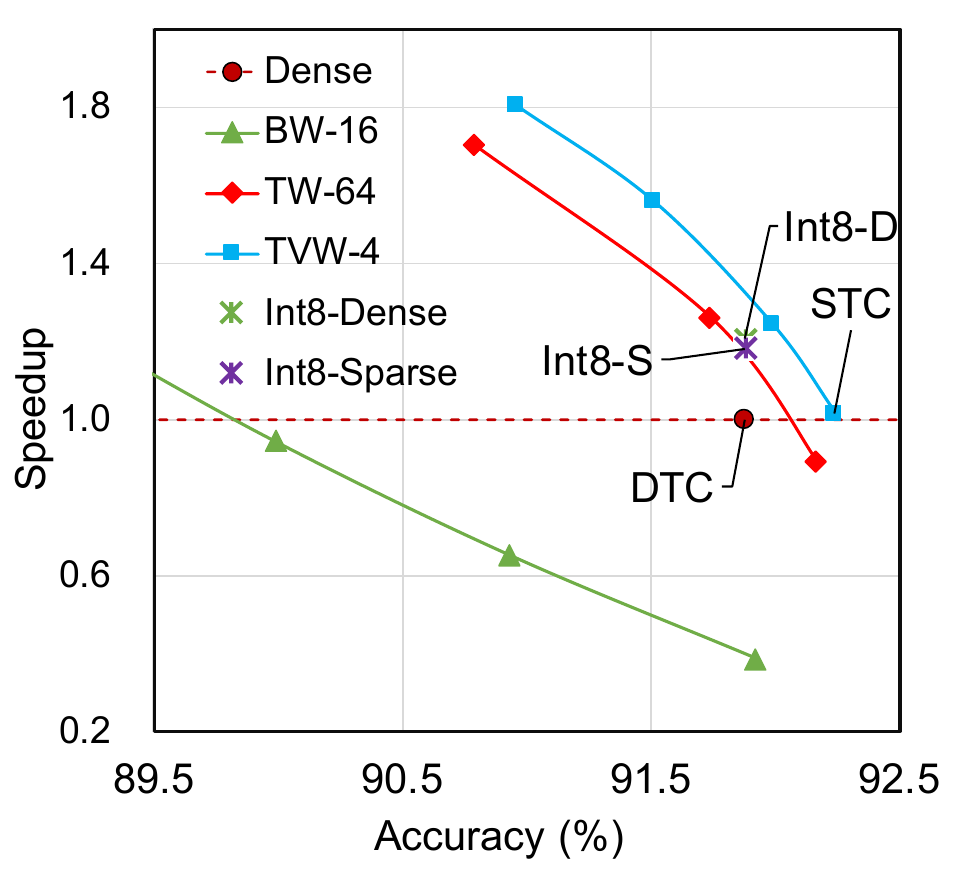}
        \vspace*{-0.7cm}
        \caption{VGG-16}
        \label{fig:vgg_spd}
        \end{subfigure}~
    \begin{subfigure}{0.33\textwidth}
        \includegraphics[trim=0 0 0 0, clip,  width=\linewidth]{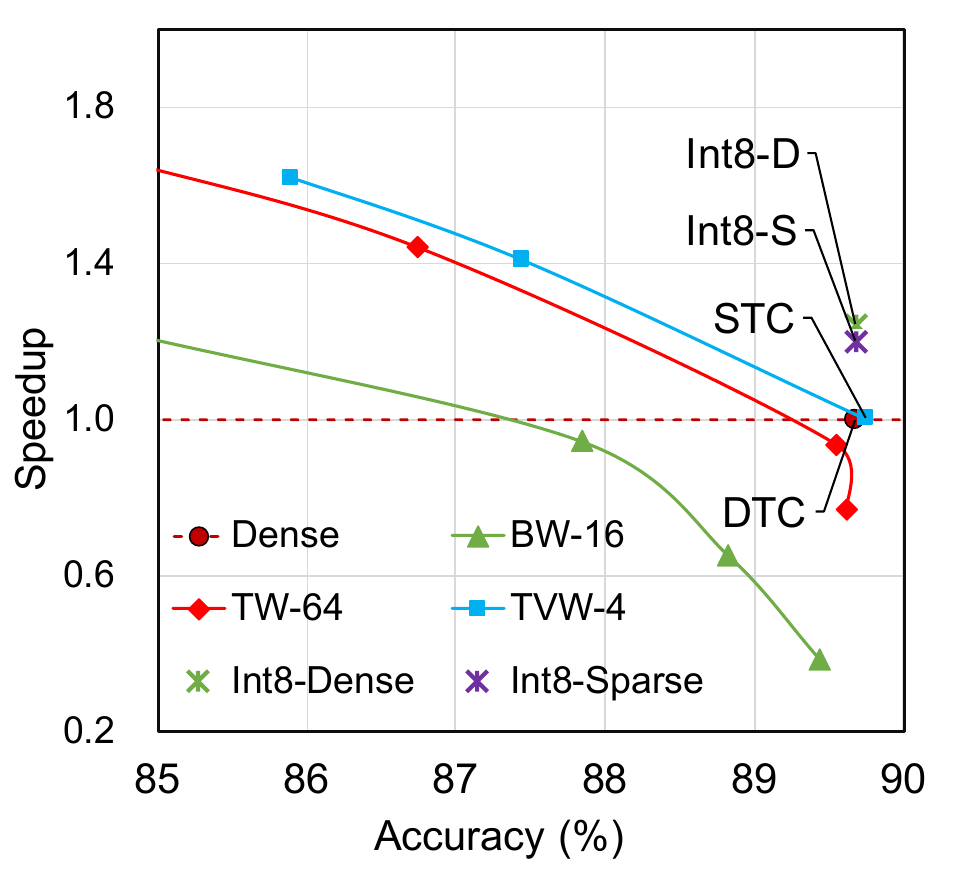}
        \vspace*{-0.7cm}
        \caption{ResNet-18}
        \label{fig:resnet18_spd}
    \end{subfigure}~
    \begin{subfigure}{0.33\textwidth}
        \includegraphics[trim=0 0 0 0, clip,  width=\linewidth]{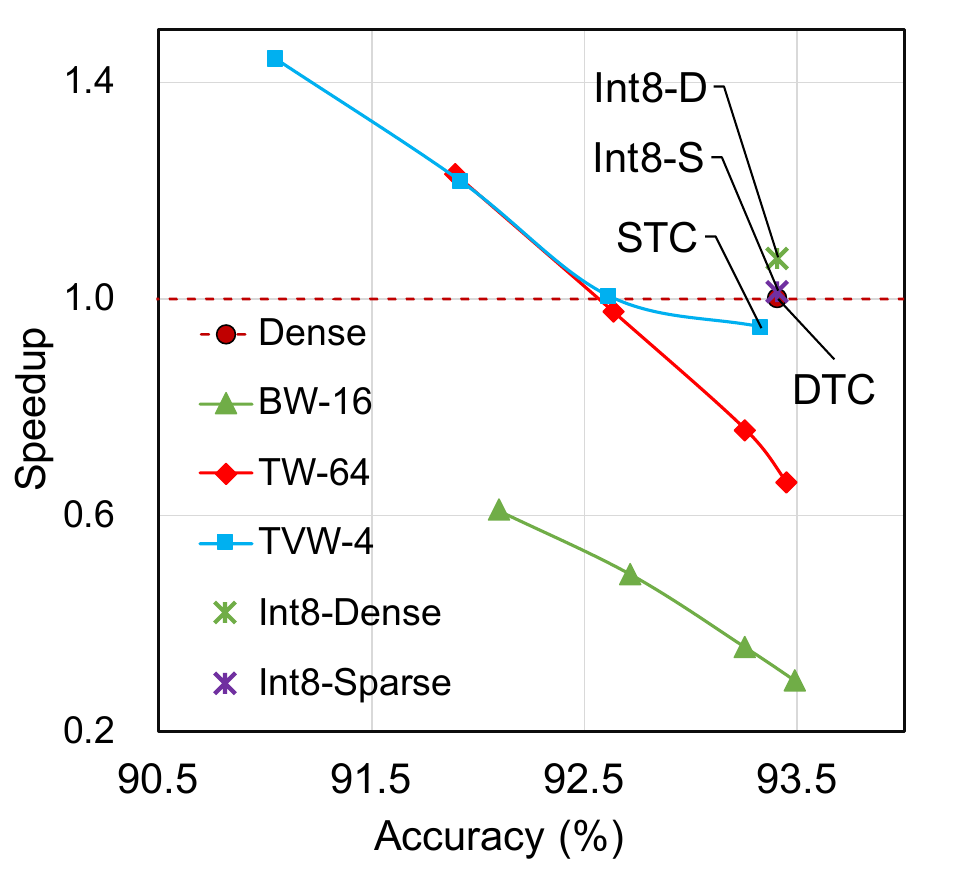}
        \vspace*{-0.7cm}
        \caption{ResNet-50}
        \label{fig:resnet50_spd}
    \end{subfigure}
        \vfill
    \begin{subfigure}{0.33\textwidth}
        \includegraphics[trim=0 0 0 0, clip, width=\linewidth]{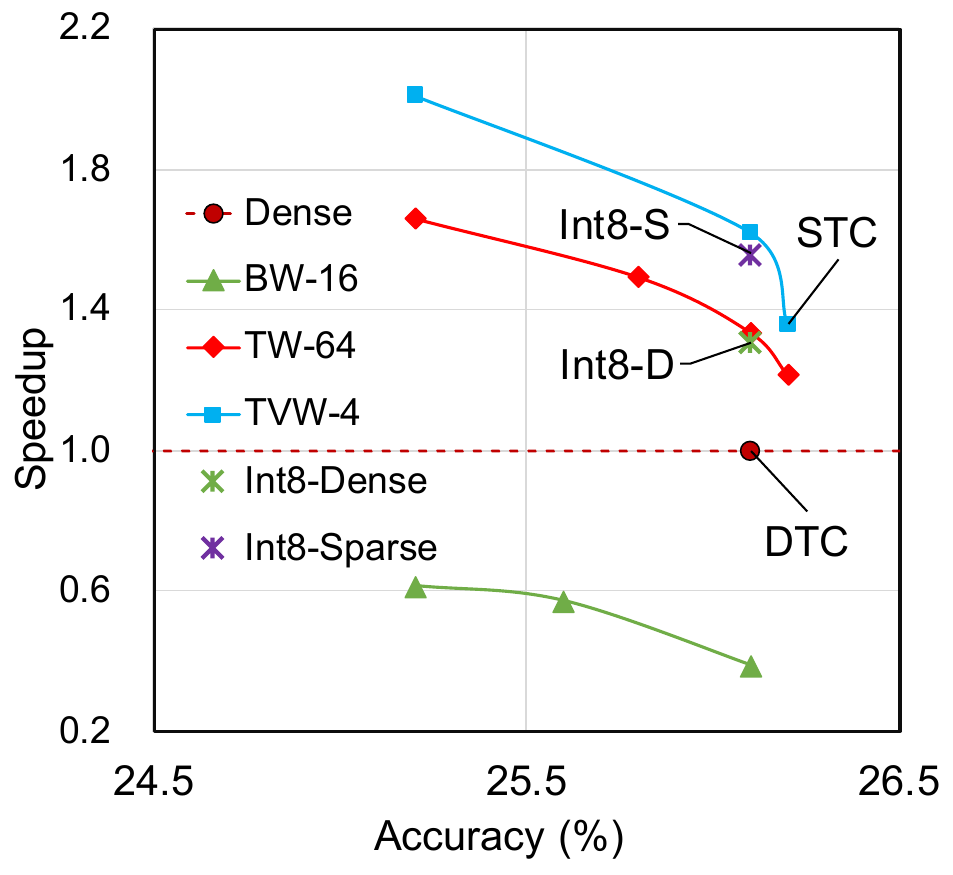}
        \vspace*{-0.7cm}
        \caption{NMT}
        \label{fig:nmt}
        \end{subfigure}~
      \begin{subfigure}{0.33\textwidth}
      \includegraphics[trim=0 0 0 0, clip, width=\linewidth]{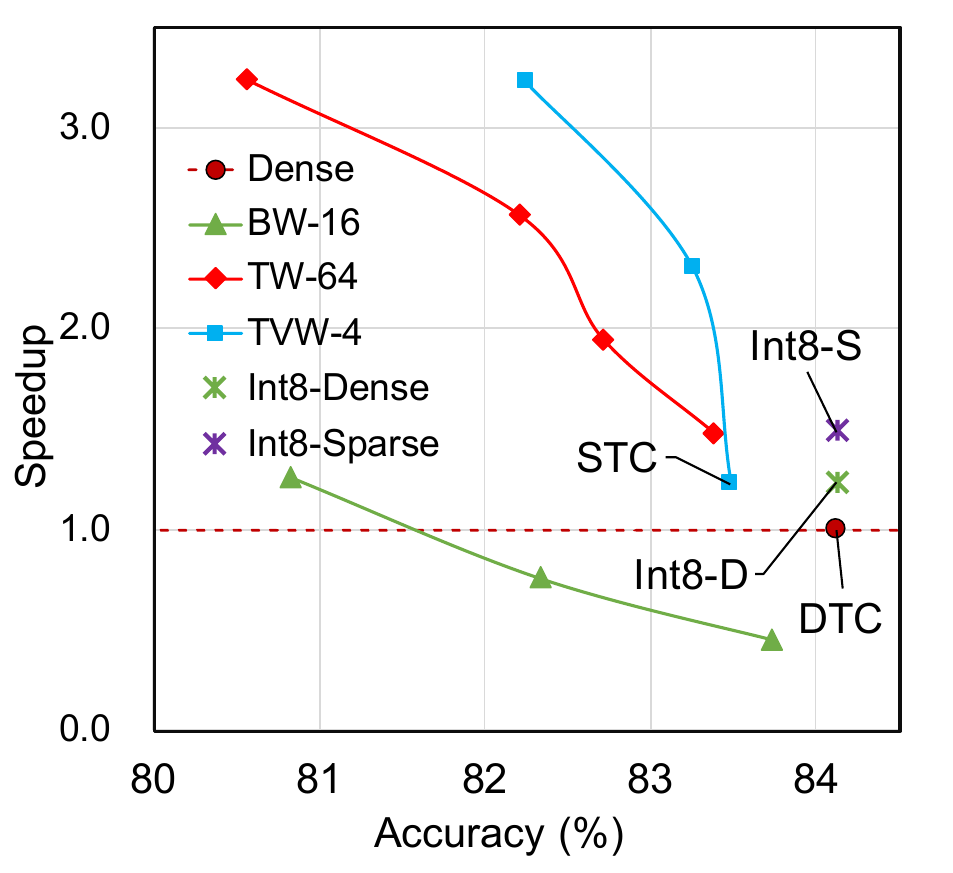}
      \vspace*{-0.7cm}
      \caption{BERT on MNLI}
      \label{fig:nmt_spd}
      \end{subfigure}~
      \begin{subfigure}{0.33\textwidth}
      \includegraphics[trim=0 0 0 0, clip,  width=\linewidth]{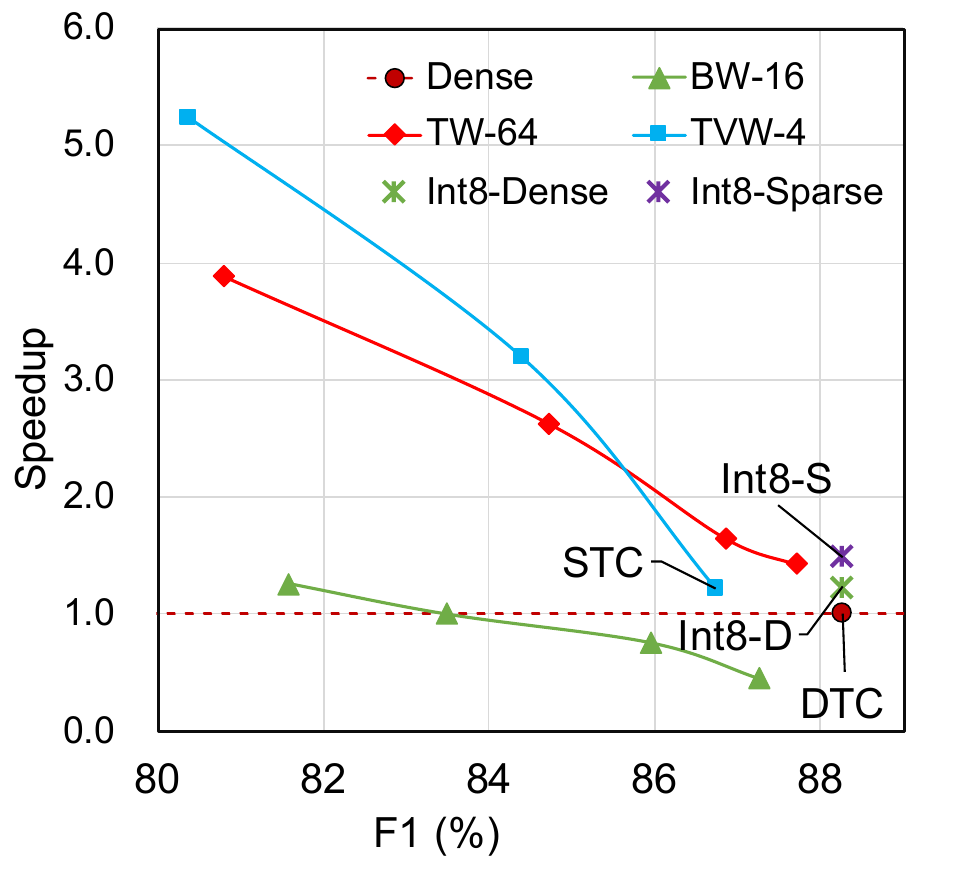}
          \vspace*{-0.7cm}
      \caption{BERT on SQuAD}
      \label{fig:squad_spd}
      \end{subfigure}

      \caption{The trade-off between latency speedup and model accuracy with GPU A100 (sparse) tensor core. 
      }
      \label{fig:speedup_all}
     \vspace*{-0.6cm}
  \end{figure*}

\paragraph{Accuracy.}
\Fig{fig:accuracy_all} shows the accuracy of different models with different pruning patterns.
The granularity of \benchmark{TW} and \benchmark{TVW} is 64 (128) for CNN (NMT and BERT) and the block size of \benchmark{BW} is $16 \times 16$, which balances the accuracy and latency speedup as our previous design space analysis suggests.
We adopt top-5 accuracy for VGG, ResNet-18, and ResNet-50.
The vector size of \benchmark{VW-16} is set to 16 as used in the original paper~\cite{zhu2019sparse}. \benchmark{VW-4} is the original sparse tensor core 2:4 sparsity on A100. Therefore, we also have the \benchmark{TVW-4} and \benchmark{TVW-16} corresponding to the \benchmark{VW-4} and \benchmark{VW-16}.

\benchmark{EW} reaches the best accuracy of all the evaluated algorithms, and \benchmark{BW} has the worst accuracy under the same sparsity except for the Resnet-50 model.
\benchmark{BW} has the largest granularity, which contains $16\times 16 = 256$ elements for each pruning. For ResNet-50, we checked the original pruning data and found that \benchmark{BW} left some smaller layers (e.g., the first layer) without pruning, surprisingly improving accuracy. 
This indicates that \benchmark{TW} can also get better accuracy if \benchmark{TW} can also skip the smaller layers or apply more comprehensive metrics and algorithms.

\benchmark{VW-16} slightly outperforms \benchmark{TW} when the sparsity is below 75\%, owing to its irregularity inside the vector with a length of 16. 
With high sparsity ($> 75\%$), \benchmark{TW} generally outperforms the \benchmark{VW}  with the exception of NMT because \benchmark{TW} has more flexibility for the high sparsity pruning and allows the uneven distribution of sparsity in a weight matrix.
\benchmark{VW}, \benchmark{BW}, and \benchmark{TW} experience a rapid accuracy drop compared to \benchmark{EW} when the sparsity is over 75\%, which suggests these models prefers irregular and unstructured sparsities.

\paragraph{Sparsity Pattern.}
\Fig{fig:pattern} shows the results of weight sparsity distributions of first layer of BERT under the $75\%$ sparsity for different patterns.
The \benchmark{EW} result shows that there exists uneven distribution across the matrix.
And \benchmark{VW} cannot fit this characteristic because it forces all prune units (vector) to have the same sparsity.
Both \benchmark{BW} and \benchmark{TW} can adapt to this sparsity locality. 
However, the granularity of \benchmark{BW} is too large and prunes complete square blocks, leading to too many blank areas and lower irregularity. 
In contrast, \benchmark{TW} only prunes columns and rows inside a tiled block, maintaining higher irregularity, as shown in \Fig{fig:pattern_d}.

\benchmark{TVW} contains the \benchmark{TW} pattern and \benchmark{VW} pattern.
Therefore, \benchmark{TVW-4} in \Fig{fig:pattern_e} is the most similar to the \benchmark{EW} with high irregularity.
\benchmark{TW} prunes the weight tensor with the uneven distribution, and \benchmark{VW} is irregular inside each vector.
Naturally, \benchmark{TVW-16} can combine the irregularity of \benchmark{VW-16} and uneven distribution of \benchmark{TW} and achieve significantly superior accuracy over the \benchmark{TW} and \benchmark{VW-16}. 
\benchmark{TVW-4} also benefits from the advantages of \benchmark{TW} and \benchmark{VW-4} and surpasses the accuracy of \benchmark{TW}.
Still, the accuracy curve of \benchmark{TVW-4} is slightly lower than the \benchmark{TVW-16} because its vector size is smaller than \benchmark{TVW-16}, and its pruning pattern is a fixed 2:4 (50\%) sparsity pattern. 
As such, the curve of \benchmark{TVW-4} can only start from 50\%, which is exactly the \benchmark{VW-4} sparsity without \benchmark{TW} pruning.
Fortunately, \benchmark{TVW-4} is supported by the existing GPU, but \benchmark{TVW-16} is not.

In summary, \benchmark{TVW} combines the advantages of \benchmark{TW} and \benchmark{VW}, and achieves the second-best accuracy after the \benchmark{EW}. This advantage of \benchmark{TVW} can provide more potential opportunities to accelerate sparse models.


\begin{figure*}[t]
    \vspace*{-.2cm}
    \centering
    \begin{subfigure}{0.33\textwidth}
        \includegraphics[trim=0 0 0 0, clip,  width=\linewidth]{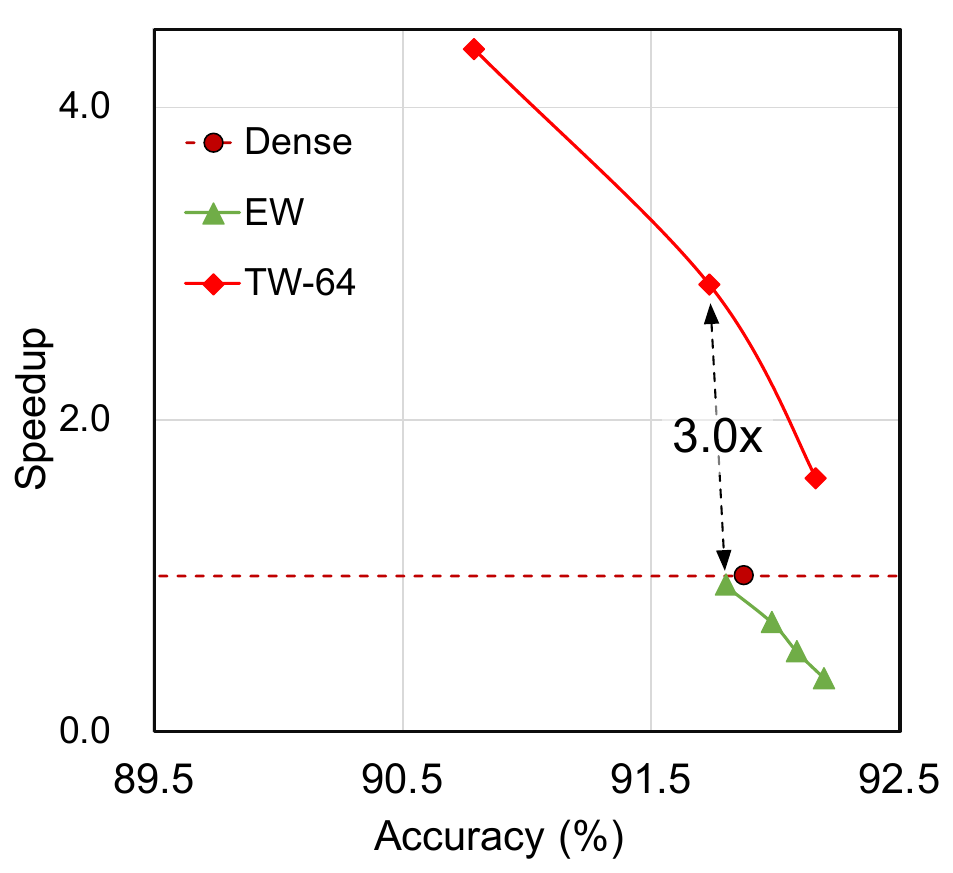}
        \vspace*{-0.7cm}
        \caption{VGG-16}
        \label{fig:vgg_spd_cuda}
        \end{subfigure}~
    \begin{subfigure}{0.33\textwidth}
        \includegraphics[trim=0 0 0 0, clip,  width=\linewidth]{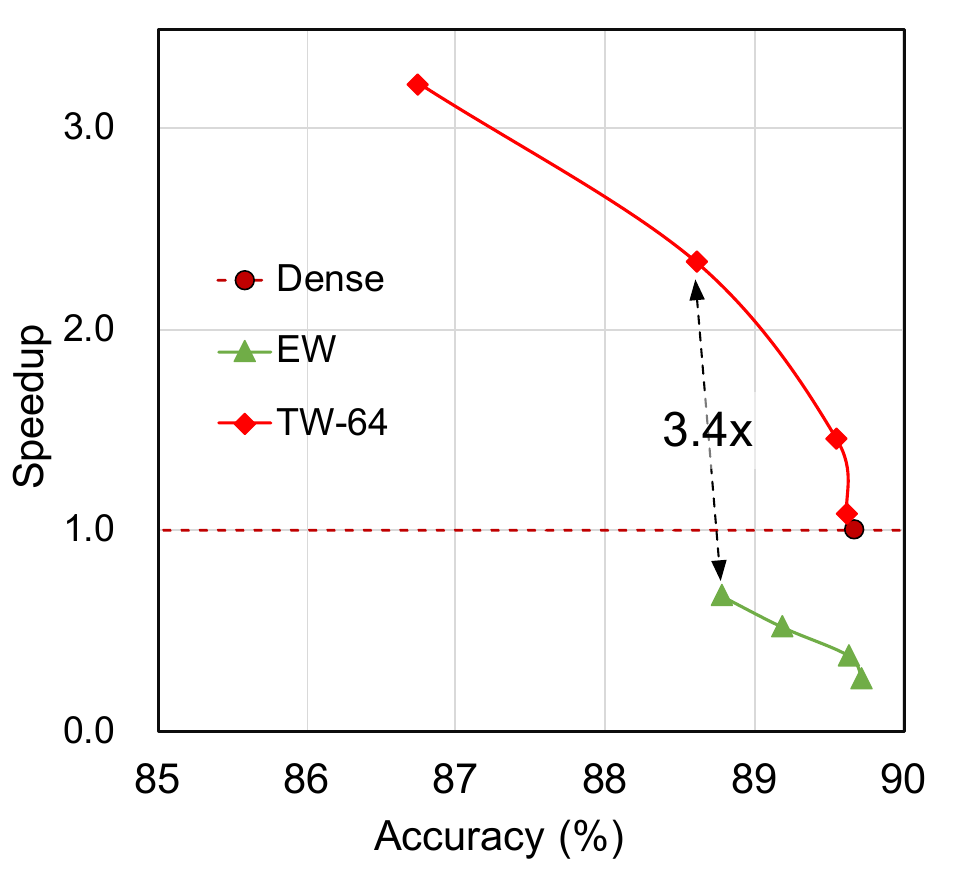}
        \vspace*{-0.7cm}
        \caption{ResNet-18}
        \label{fig:resnet18_spd_cuda}
    \end{subfigure}~
    \begin{subfigure}{0.33\textwidth}
        \includegraphics[trim=0 0 0 0, clip,  width=\linewidth]{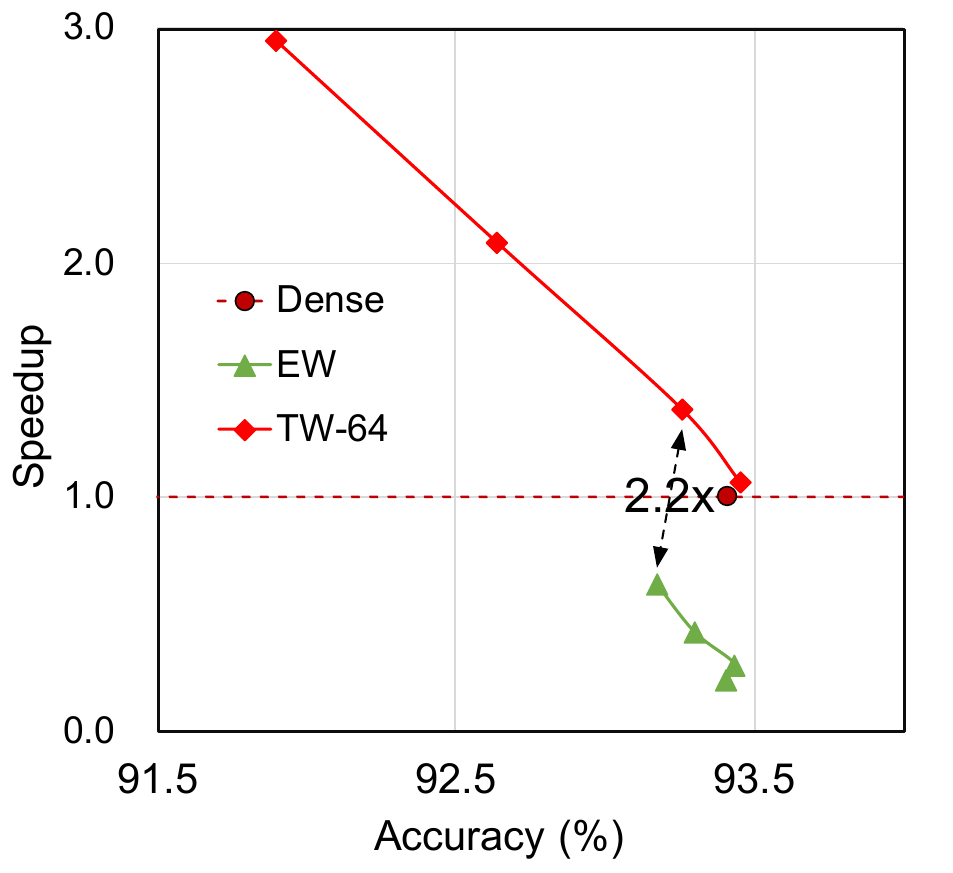}
        \vspace*{-0.7cm}
        \caption{ResNet-50}
        \label{fig:resnet50_spd_cuda}
    \end{subfigure}
        \vfill
    \begin{subfigure}{0.33\textwidth}
        \includegraphics[trim=0 0 0 0, clip, width=\linewidth]{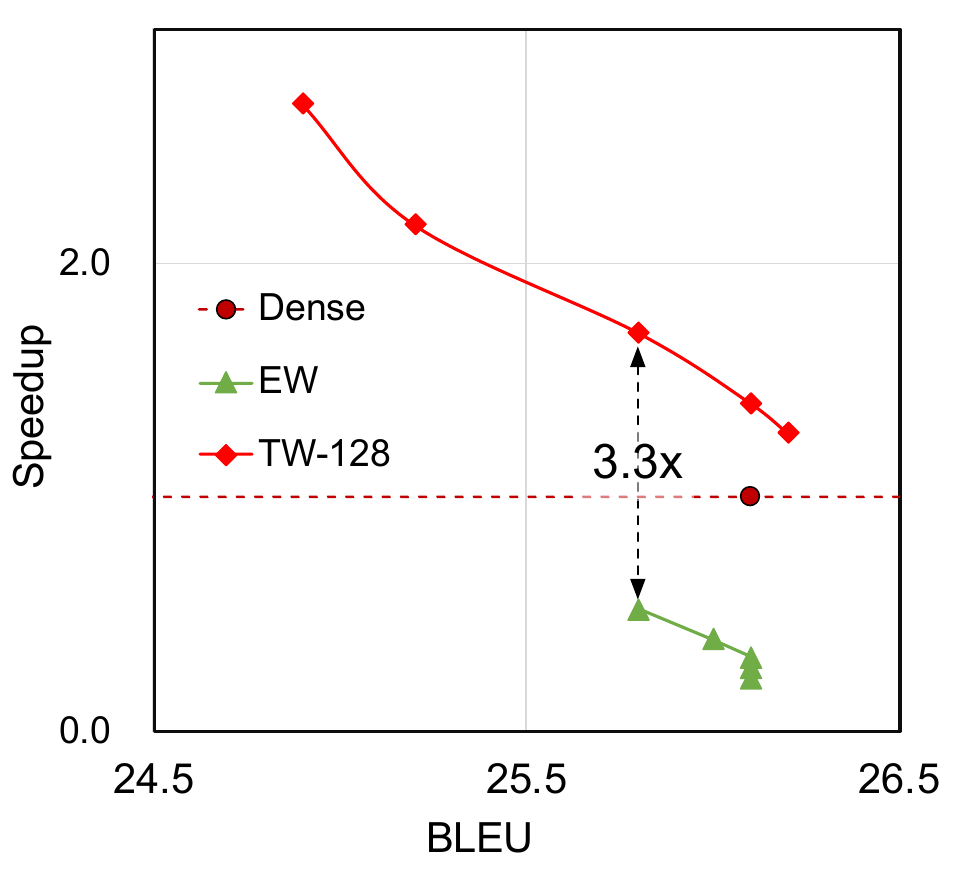}
        \vspace*{-0.7cm}
        \caption{NMT}
        \label{fig:nmt}
        \end{subfigure}~
      \begin{subfigure}{0.33\textwidth}
      \includegraphics[trim=0 0 0 0, clip, width=\linewidth]{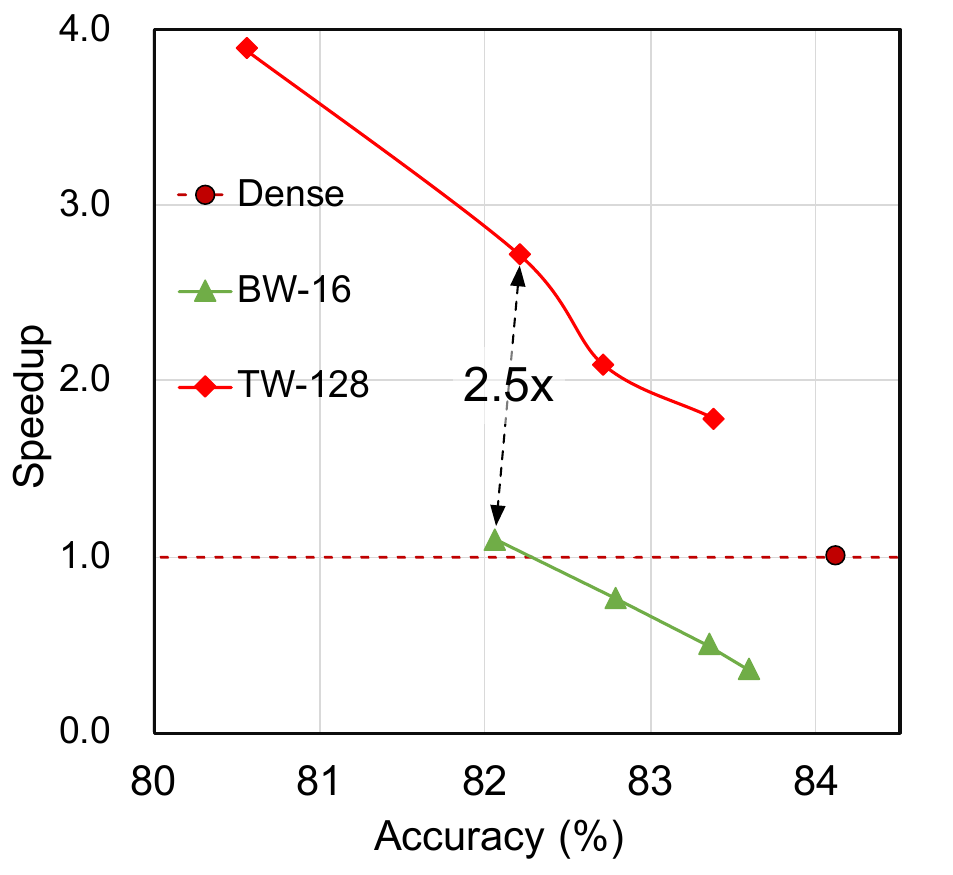}
      \vspace*{-0.7cm}
      \caption{BERT on MNLI}
      \label{fig:nmt_spd_cuda}
      \end{subfigure}~
      \begin{subfigure}{0.33\textwidth}
      \includegraphics[trim=0 0 0 0, clip,  width=\linewidth]{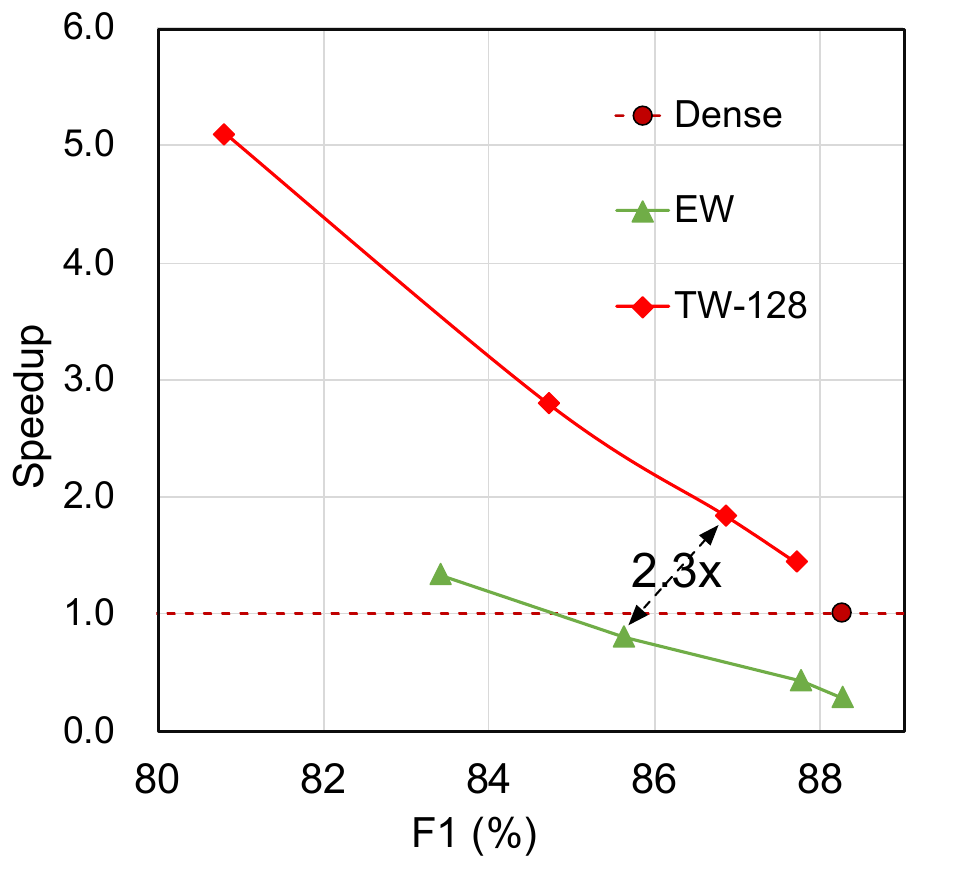}
          \vspace*{-0.7cm}
      \caption{BERT on SQuAD}
      \label{fig:squad_spd_cuda}
      \end{subfigure}

      \caption{The trade-off between latency speedup and model accuracy with GPU A100 CUDA core. }
      \label{fig:speedup_all_cuda}
     \vspace*{-0.6cm}
  \end{figure*}

  \subsection{Speedup vs. Accuracy Comparison}
\Fig{fig:speedup_all} and \Fig{fig:speedup_all_cuda} compare the trade-off of latency speedup and model accuracy based on \benchmark{TW} and other patterns including \benchmark{BW}, \benchmark{VW} on A100 (sparse) tensor core and \benchmark{EW} on the CUDA core. 
We compare the \benchmark{TVW}, \benchmark{TW}, \benchmark{VW}, and \benchmark{BW} running on the (sparse) tensor core, and compare the \benchmark{TW} and \benchmark{EW} on the CUDA core. 
The granularity $G$ of \benchmark{TW} and \benchmark{TVW} is 64 for CNNs and 128 for BERT and NMT, the granularity of \benchmark{BW} is 16, and \benchmark{TVW-4} employs \benchmark{VW-4}, which is exactly the original sparse tensor core 2:4 pattern. 
As such, the starting point (STC) of the \benchmark{TVW-4} curve is the \benchmark{VW-4} results.
\benchmark{EW} can only run on CUDA core with cuSparse. 
The speedup is calculated against dense models on the tensor core and CUDA core separately. 
The experimental results demonstrate that \benchmark{TVW} and \benchmark{TW} can extend the latency-accuracy Pareto frontier on tensor cores. In contrast, other sparsity patterns lead to both longer latency and lower accuracy than the dense model.
Finally, we compare the latency speedup of various patterns with the same level of accuracy drop (BERT with $< 2\%$ accuracy or F1 drop, VGG, ResNet-18, and ResNet-50 with $< 2\%$ top-5 accuracy drop, and NMT with $< 1$ BLEU drop).

\paragraph{Tensor Core.}
For the tensor core results in \Fig{fig:speedup_all}, \benchmark{TVW} achieves an average speedup of \textbf{$\mathbf{1.85\times}$}, and \benchmark{TW} gets a \textbf{$\mathbf{1.70\times}$} speedup over the original dense GEMM.
\benchmark{TVW} performs best in most scenarios except for BERT on the SQuAD dataset and ResNet-50.
We find that the SQuAD dataset is sensitive to sparsity. However, the minimum sparsity of \benchmark{TVW} is the fixed 50\% because of the hardware constraint from the sparse tensor core.
Fortunately, \benchmark{TW} complements this flaw of \benchmark{TVW} at lower sparsity still with considerable speedup.
For ResNet-50, \benchmark{TW} and \benchmark{TVW} have almost the same trends after 50\% sparsity because they also have similar accuracy trends in \Fig{fig:resnet50_acc}.

From the perspective of \benchmark{VW-4} (i.e., the STC point), \benchmark{VW-4} achieves an average of $\mathbf{1.25\times}$ speedup for the NMT and BERT on MNLI and SQuAD, but there is no  ($\mathbf{0.98\times}$) speedup for CNN models (VGG, ResNet-18, and ResNet-50), whose shape of GEMM computation is smaller than Transformer-based (BERT) and NMT models.
In addition, compared with the experiment of the large GEMM in \Fig{fig:g_latency_t}, \benchmark{VW-4} gets $\mathbf{1.67\times}$ speedup on the shape of ($4096 \times 4096 \times 4096$).
Evidently, \benchmark{VW} is signiﬁcantly affected by the shape of GEMM computation. In some corner cases, \benchmark{VW} performs badly due to the inefﬁcient execution of GEMM with a small shape.
Combined with \benchmark{TW}, \benchmark{TVW} can be extended to a higher and more flexibale sparsity over \benchmark{VW}-only pattern and achieve meaningful speedup for DNN models.
Even though \benchmark{BW} achieves better accuracy in some models (e.g., ResNet-50 in \Fig{fig:resnet50_acc}), \benchmark{TVW} still surpasses \benchmark{BW} by $\mathbf{2.75\times}$ because \benchmark{TVW} adopts larger tiling size ($64$) but smaller granularity ($1\times 64$) for the tiled GEMM algorithm with high efficiency to run on GPU tensor core.

\paragraph{Int8 Quantization.}
We also compare the \benchmark{Int8} quantization on the (sparse) tensor core. 
Based on our survey~\cite{wu2020integer}, the \benchmark{Int8} quantization exhibits almost no accuracy loss across all models. As for \benchmark{Int8-Sparse}, we consider them as a reference in \Fig{fig:speedup_all}, given the absence of accuracy reports.
We find that the \benchmark{Int8-Sparse} has the same trends as the \benchmark{VW-4} (FP16) sparse pattern because they have a similar implementation on A100 GPU with the CUTLASS library.
Due to the same reason, \benchmark{Int8-Sparse} performs worse than \benchmark{Int8-Dense} on CNN models, which have smaller GEMM shapes that are not friendly to the \benchmark{VW} sparsity. For the Transformer-based models, \benchmark{Int8} can achieve meaningful speedups over FP16 dense models. \benchmark{Int8-Dense} achieves $1.26\times$, and \benchmark{Int8-Sparse} has $1.51\times$ speedups.
The speedup is fixed and marginal compared to the pruning-based \benchmark{TW} and \benchmark{TVW}, which can have higher performance and more flexibility.

\paragraph{CUDA Core.}
For the CUDA core results in \Fig{fig:speedup_all_cuda}, \proj{} achieves an average speedup of $\mathbf{2.43 \times}$ over the dense GEMM with the minor accuracy loss.
In most models, \benchmark{EW} cannot deliver meaningful speedups because \benchmark{EW} is an unstructured pattern, leading large amount of irregular memory accesses.
As such, \proj{} is $\mathbf{2.78\times}$ faster than \benchmark{EW} with similar accuracy loss.
Please notice that \benchmark{EW} can only run on the CUDA core.
Tensor core-based implementation performs a significant advantage (about $10\times$ speedup presented in \Fig{fig:g_latency_c}) over CUDA core.
Therefore, when the accuracy of \benchmark{TVW} is similar to the accuracy of \benchmark{EW},
\benchmark{TVW} outperforms $\mathbf{22.18\times}$ speedup over \benchmark{EW}.

In summary, \benchmark{TVW} and \benchmark{TW} achieve meaningful latency reduction on GPU due to their compatibility with dense GEMM, while all other sparsity patterns cause the slowdown. The design of \benchmark{TVW} with more flexible sparsity can complement other sparsity patterns.  That makes \benchmark{TVW} can accommodate some resource-constraint scenarios, such as mobile systems.

\section{Conclusion}\label{sec:conclude}

In this work, we propose co-designing the tiling of matrix multiplication and DNN model pruning pattern to balance the irregularity for the model accuracy and compatibility for dense GEMM computation.
We study an efficient software-only implementation of our proposed sparsity pattern, \proj, that leverages the GPU's tensor core accelerator and concurrency features.
We further exploit the characteristic of the latest GPU A100 and design a more flexible sparsity pattern \benchmark{TVW} combining the advantages from \benchmark{VW} and \benchmark{TW}.
We demonstrate its capability of model accuracy preserving and high performance speedup on the state-of-the-art DNN models.
Finally, \benchmark{TVW} achieves significant $2.75\times$ and $22.18\times$ speedups over block sparsity and unstructured sparsity.

\section*{Acknowledgments}
This work was supported by the National Key R\&D Program of China under Grant 2021ZD0110104, the National Natural Science Foundation of China (NSFC) grant (62222210, U21B2017, and 62072297).
The authors would like to thank the anonymous reviewers for their constructive feedback for improving the work. 
Any opinions, findings, and conclusions in this paper are those of the authors only and do not necessarily reflect the views of our sponsors.

\bibliographystyle{IEEEtran}
\bibliography{references}


\section{Biography Section}
\begin{IEEEbiography}[\vspace{-20pt} {\includegraphics[width=1in,height=1.25in,clip,keepaspectratio]{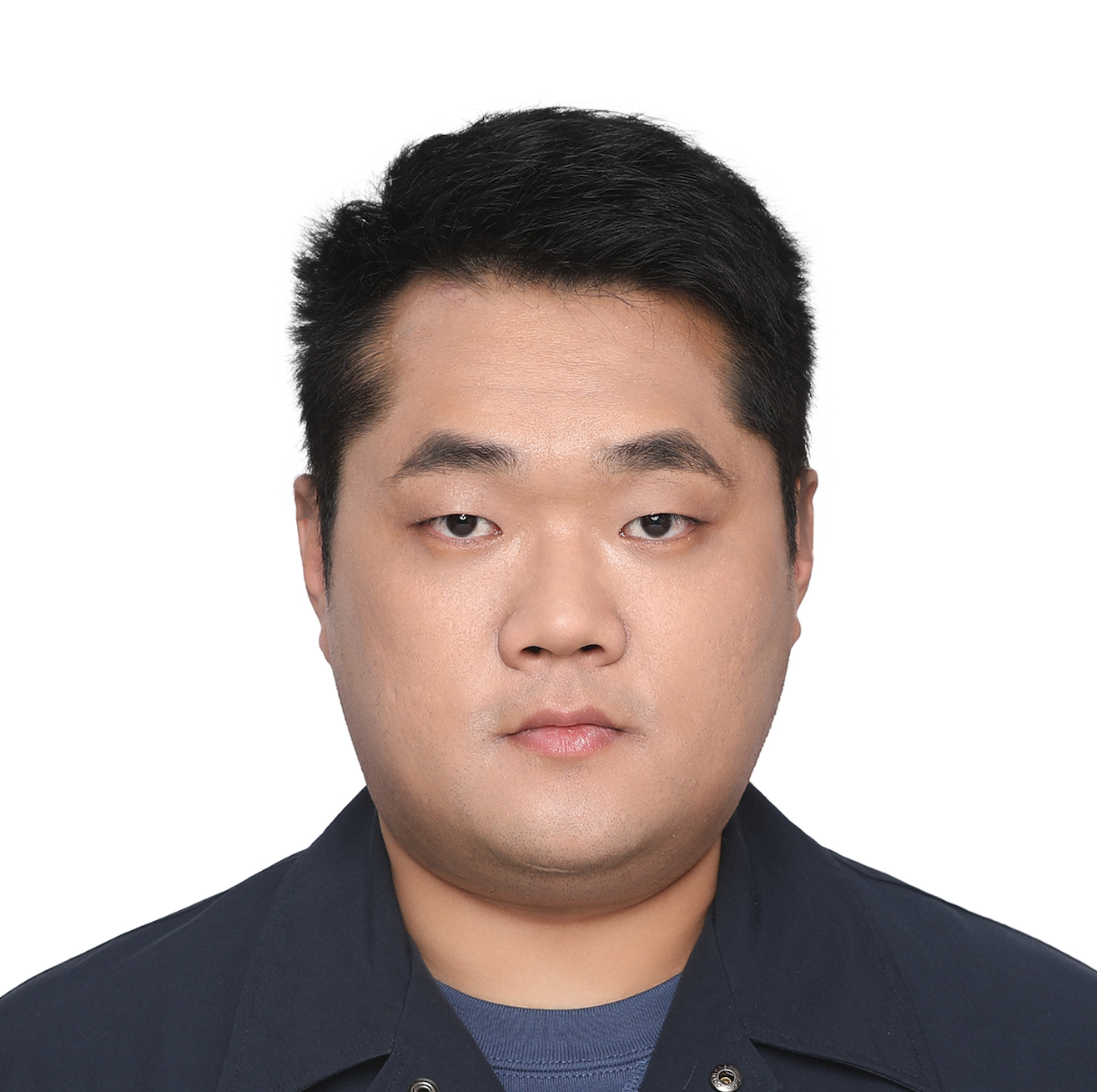}}]
        {Cong Guo} received his B.Sc. degree from Shenzhen University, China. He is currently a Ph.D. candidate in computer science under the supervision of Dr. Jingwen Leng at the Department of Computer Science and Engineering of Shanghai Jiao Tong University, China. His research interests include computer architecture, high-performance computing, and AI accelerator design.
\end{IEEEbiography}
\vspace{-20pt}
\begin{IEEEbiography}[\vspace{-30pt} {\includegraphics[width=0.9in,height=1.2in,clip,keepaspectratio]{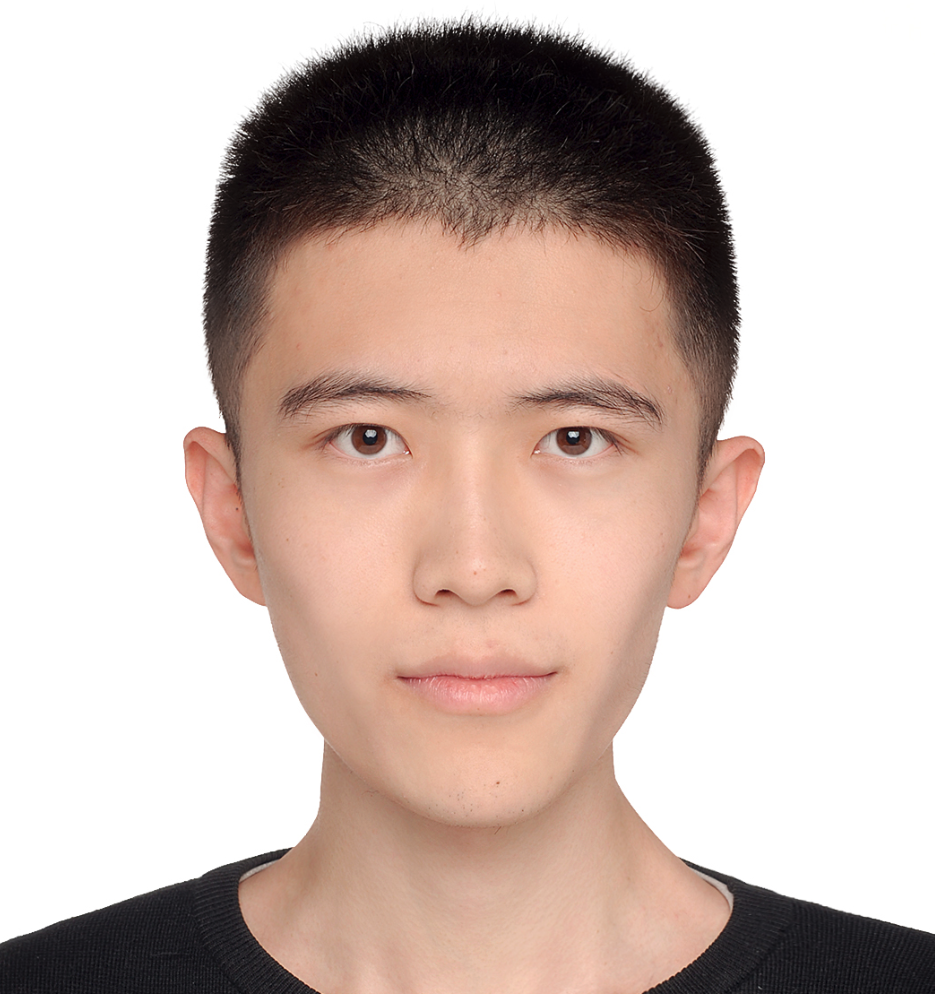}}]
        {Fengchen Xue} received the bachelor's degree from the Department of Computer Science and Engineering, Shanghai Jiao Tong University, China. He is studying for a master's degree at Zhejiang University. His research interests include high performance computing, machine learning system and ray-tracing hardware.
\end{IEEEbiography}
\vspace{-20pt}
\begin{IEEEbiography}[\vspace{-15pt} {\includegraphics[width=1in,height=1.25in,clip,keepaspectratio]{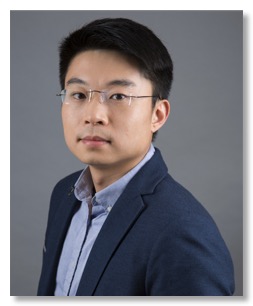}}]
        {Jingwen Leng} is a tenured Associate Professor in John Hopcroft Computer Science Center and Computer Science and Engineering Department at Shanghai Jiao Tong University. He received the Ph.D. degree from the University of Texas at Austin. He was the lead co-author for GPUWattch, one of the most widely used open-sourced GPU power model. He is currently interested at building intelligent and robust system for artificial intelligence.
\end{IEEEbiography}
\vspace{-20pt}
\begin{IEEEbiography}[\vspace{-30pt} {\includegraphics[width=0.9in,height=1.25in,clip,keepaspectratio]{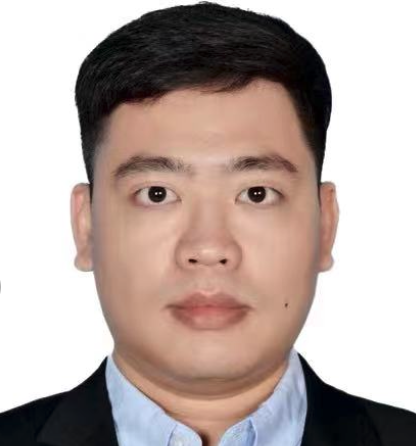}}]
        {Yuxian Qiu} received the PhD degree from the Department of Computer Science and Engineering, Shanghai Jiao Tong University, China, March 2023. He works in the TensorRT team at NVIDIA. He is interested in building interpretable and robust deep learning systems.
\end{IEEEbiography}
\vspace{-20pt}
\begin{IEEEbiography}[\vspace{-30pt} {\includegraphics[width=1in,height=1.25in,clip,keepaspectratio]{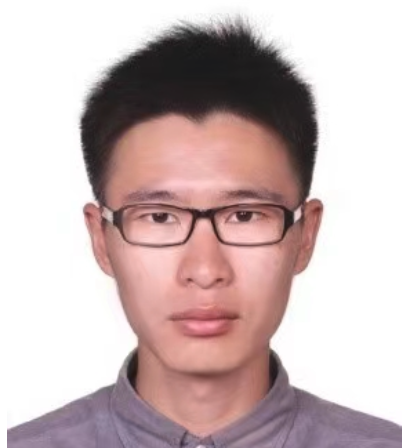}}]
        {Yue Guan} is currently a PhD candidate at the Department of Computer Science and Engineering, Shanghai Jiao Tong University, China. He received his master degree from Graduate School of Information, Production and Systems, Waseda University. His research interests include efficient deep learning systems, algorithms and their co-design.
\end{IEEEbiography}
\vspace{-20pt}

\begin{IEEEbiography}[\vspace{-30pt}{\includegraphics[width=1in,height=1.25in,clip,keepaspectratio]{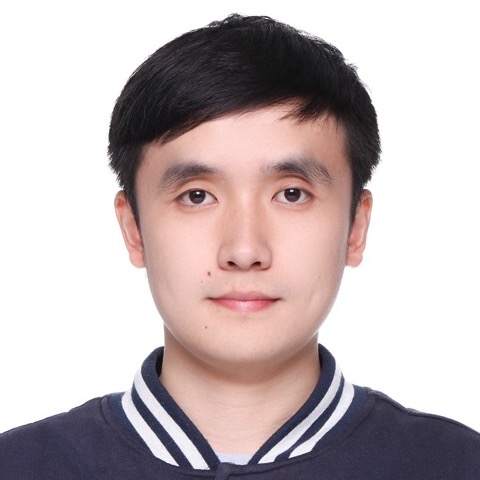}}]
        {Weihao Cui} received his B.Sc. degree from Shanghai Jiao Tong University, China. He is currently a Ph.D. candidate in the field of computer science under the supervision of Dr. Quan Chen in Department of Computer Engineering Faculty of Shanghai Jiao Tong University, China. His research interests include high-performance computing and resource management of accelerators in datacenters.
\end{IEEEbiography}
\vspace{-20pt}
\begin{IEEEbiography}[\vspace{-20pt}{\includegraphics[width=1in,height=1.25in,clip,keepaspectratio]{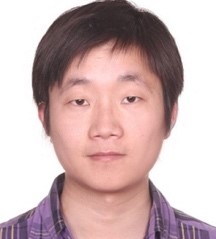}}]
        {Quan Chen} is a professor in the Department of Computer Science and Engineering, Shanghai Jiao Tong University, China. His research interests include High performance computing, Task Scheduling in various architectures, Resource management in Datacenter, Runtime System and Operating System. He got his Ph.D. degree at June 2014 from the Department of Computer Science and Engineering, Shanghai Jiao Tong University, China.
\end{IEEEbiography}
\vspace{-20pt}
\begin{IEEEbiography}[\vspace{-10pt}{\includegraphics[width=1in,height=1.25in,clip,keepaspectratio]{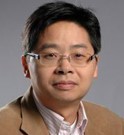}}]
        {Minyi Guo} (Fellow, IEEE) received the Ph.D. degree in computer science from the University of Tsukuba, Japan. He is currently Zhiyuan Chair professor in the Department of Computer Science and Engineering, Shanghai Jiao Tong University, China. His present research interests include parallel/distributed computing, compiler optimizations, embedded systems, pervasive computing, big data and cloud computing. He is now on the editorial board of IEEE Transactions on Parallel and Distributed Systems, IEEE Transactions on Cloud Computing and Journal of Parallel and Distributed Computing. Dr. Guo is a fellow of IEEE, and a fellow of CCF.
\end{IEEEbiography}

\end{document}